\pdfoutput=1
\documentclass[11pt]{article}
\usepackage{amsmath}
\usepackage{amssymb}
\usepackage{slashed}
\usepackage{graphicx,epstopdf}
\usepackage{cite}
\usepackage{pbox}
\usepackage{geometry}
\usepackage{float}
\usepackage{array}
\usepackage{amsfonts}
\usepackage{graphicx}
\usepackage[font={small}]{caption}
\usepackage[table]{xcolor}
\usepackage{xcolor}
\usepackage{hyperref}
\hypersetup{
    colorlinks=true,
    linkcolor=red,
    filecolor=magenta,
    urlcolor=blue,
   citecolor=blue,
}

\definecolor{RawSienna}{cmyk}{0,0.72,1,0.45}
\definecolor{dgreen}{rgb}{0.0,0.42,0.13}
\definecolor{darkblue}{rgb}{0.0, 0.0, 0.55}
\definecolor{cornellred}{rgb}{0.7, 0.11, 0.11}
\definecolor{calpolypomonagreen}{rgb}{0.08, 0.5, 0.5}

\def\beq{\begin{equation}}
\def\eeq{\end{equation}}
\def\bea{\begin{eqnarray}}
\def\eea{\end{eqnarray}}
\makeatletter
\@addtoreset{equation}{section}

\begin{document}
\title{\LARGE \bf CP transformed mixed $\mu\tau$ antisymmetry for neutrinos and its consequences}
\author{{\bf Roopam Sinha$^1$\footnote{roopam.sinha@saha.ac.in}, Probir Roy$^2$\footnote{probirrana@gmail.com}, Ambar Ghosal$^1$\footnote{ambar.ghosal@saha.ac.in},}\\
1. Astroparticle Physics and Cosmology Division\\Saha Institute of Nuclear Physics, HBNI, Kolkata 700064, India\\2. Center for Astroparticle Physics and Space Science \\ Bose Institute, Kolkata 700091, India}

\maketitle

\begin{abstract}
We propose a complex extension of mixed $\mu\tau$ antisymmetry in the neutrino Majorana mass matrix $M_\nu$. This can be implemented in the Lagrangian by a generalized CP transformation (labeled by a mixing parameter $\theta$) on the left-chiral flavor neutrino fields. We investigate its implications for leptonic CP violation and neutrino phenomenology in general. Interestingly, the $\mu\tau$ mixing parameter $\theta$ gets correlated with the Dirac CP phase $\delta$ and the atmospheric mixing angle $\theta_{23}$ through an analytical relation. In general, for arbitrary $\theta$, both $\theta_{23}$ and $\delta$ are nonmaximal. We discuss the corresponding results for the CP asymmetry parameter $A_{\mu e}$ in neutrino oscillation experiments. For a nonmaximal $\delta$, one of the two Majorana phases is different from $0$ or $\pi$, thereby leading to nonvanishing Majorana CP violation with observable consequences for the neutrinoless double beta ($\beta\beta0\nu$) decay process. We numerically work out in detail the predictions for that process in relation to various ongoing and forthcoming experiments. We also work out the predictions of our scheme on flavor flux ratios at neutrino telescopes. While exact CP transformed $\mu\tau$ interchange antisymmetry  ($\theta=\pi/2$) leads to an exact equality among those ratios, taking a value $0.5$, a tiny deviation can cause a drastic change in them. Careful measurement of these flux ratios in future will further constrain the parameter $\theta$.
\end{abstract}


\section{Introduction} The theoretical origin of the masses, mixing pattern and CP properties\cite{King:2015aea} of the three light neutrinos continues to be an unsettled story. Experimentally, the three mixing angles and the two mass-squared differences are already known to a reasonably good accuracy while a fairly tight cosmological upper bound\cite{Aghanim:2016yuo} of $0.17$ eV exists on the sum of the three masses. The solar mixing angle $\theta_{12}$ is close to $33.62^\circ$ while the reactor mixing angle $\theta_{13}$ is known to be largely nonzero and approximately equal to $8.5^\circ$. The latest update on a combined global fit of neutrino oscillation data from experiments such as T2K\cite{Abe:2017bay}, NO$\nu$A\cite{Adamson:2017qqn}, MINOS\cite{minos}, RENO\cite{reno} prefers a near maximal value of atmospheric mixing angle $\theta_{23}$. Although the octant of $\theta_{23}$ is yet unknown the best fit values are $\theta_{23}=47.2^\circ$ for NO and $\theta_{23}=48.1^\circ$ for IO, preferring the higher octant for both. For the Dirac CP phase $\delta$, the current best fit values are close to $234^\circ$ for NO and $278^\circ$ for IO. Its CP conserving values (i.e., $\delta=0,\pi$) are allowed at slightly above $1\sigma$ and $\delta=\pi/2$ is disfavoured at 99\% CL\cite{Esteban:2016qun} In this scenario, $\delta=3\pi/2$ and any slight deviation from it are still permitted as interesting possibilities. In so far as the precision measurements of $\delta$ and $\theta_{23}$ are concerned, we are at a decisive moment in time since the neutrino mass models that survive current phenomenological constraints and predict a co-bimaximal mixing ($\theta_{23}=\pi/4, \delta=\pi/2$ or $3\pi/2$)\cite{Grimus:2003yn} would be subjected to a stringent experimental test. Also, the nature of the light neutrinos, whether Dirac or Majorana, remains shrouded in mystery. Perhaps future experiments will resolve the matter through a signature of the neutrinoless double $\beta-$decay process which crucially depends upon the values of the two Majorana phases of the neutrinos.

Let us first consider various discrete flavor symmetries in the $\mu\tau$ sector of neutrinos which have been proposed to understand the observed pattern of neutrino mixing. One class of such symmetries entails $\mu\tau$ mixing\cite{Samanta:2018efa}, to wit an invariance under the transformation \begin{equation}
\nu_{Ll}\to G_{lm}^\theta \nu_{L m}.\label{ram}
\end{equation} Here $G^{\theta}$ is a generator of a residual $\mathbb{Z}_2$ symmetry effecting the mixing, $l,m$ span the flavor indices $e,\mu,\tau$ while the subscript $L$ denotes left-chiral flavor neutrino fields. In neutrino flavor space $G^\theta$ has the generic form \begin{equation}G^\theta=
\begin{pmatrix}-1 & 0 & 0\\0 & -\cos\theta & \sin\theta\\
0 & \sin\theta & \cos\theta\end{pmatrix}\label{chat},
\end{equation} where $\theta$ is a mixing parameter. The location of the minus sign in \eqref{chat} is because of our convention of choosing $\det G^\theta$ to be $+1$ without any loss of generality. The special case of \eqref{ram} for $\theta=\pi/2$ has been known in the literature as $\mu\tau$ interchange symmetry which can stem from some high energy flavor symmetry group such as $S_4$\cite{Ishimori:2010au}. Further, a substantial body of work\cite{mutaus} exists investigating the phenomenological consequences of \eqref{ram}. It has been found that the reactor mixing angle $\theta_{13}$ vanishes if one imposes the symmetry \eqref{ram} with \eqref{chat}. Since this possibility has now been excluded at more than $10\sigma$\cite{An:2015rpe}, this symmetry has to be discarded.

An interesting variant of \eqref{ram} is the symmetry of CP transformed\cite{CPt} $\mu\tau$ mixing, as proposed in Ref.\cite{Chen:2015siy}. This is an invariance of the neutrino Majorana mass term under the transformation \begin{equation}
\nu_{Ll}\to iG^\theta_{lm}\gamma^0\nu^C_{Lm}\label{paf}
\end{equation} with $G^\theta$ as in \eqref{chat} and $\nu_{Ll}^C=C\overline{(\nu_{Ll})}^T$. The corresponding phenomenological consequences have been studied\cite{Chen:2015siy}. A different approach using the idea of littlest $\mu\tau$ seesaw\cite{newadd} has also been recently proposed allowing slight deviations from maximal $\theta_{23}$ and maximal Dirac CP violation. It should be noted that the $\theta\to\pi/2$ limit of \eqref{paf}, referred to as a CP transformed $\mu\tau$ interchange symmetry (${\rm CP^{\mu\tau}}$), had earlier been extensively studied\cite{Grimus:2003yn} and avoids the problem of a vanishing reactor angle. However, it predicts maximal values for the atmospheric mixing angle $\theta_{23}$ and the Dirac CP phase $\delta$, namely $\theta_{23}=\pi/4$ and $\cos\delta=0$. Such a possibility, though still allowed by current experimental limits, is being challenged by ongoing and forthcoming precision measurements of these quantities. In case the maximality of either quantity is ruled out in future, CP transformed $\mu\tau$ interchange symmetry will be excluded.

In this paper, we propose a complex antisymmetric extension of \eqref{paf} using a $\mathbb{Z}_4$ generator $\mathcal{G}^{\theta}=iG^{\theta}$
\begin{equation}
\nu_{Ll}\to i\mathcal{G}^\theta_{lm}\gamma^0\nu^C_{Lm}\label{alum}.
\end{equation} A special case of such an invariance with $\theta=\pi/2$ was proposed by some of us in Ref.\cite{Samanta:2017kce}. The latter avoids the problem of a vanishing $\theta_{13}$ but leads to maximal values of the atmospheric mixing angle $\theta_{23}$ and the Dirac CP phase $\delta$. As explained above, these results may not survive for much longer. In this situation our proposal of an invariance under \eqref{alum} with $\theta\neq\pi/2$ assumes a special significance since it allows any arbitrary nonzero value of $\theta_{13}$ and nonmaximal $\theta_{23}$ depending on the parameter $\theta$. Since in this work we concentrate on the low-energy phenomenological consequences, we start from the effective field transformation \eqref{alum} without providing a larger symmetry that embeds it. In case of CP combined with a flavor symmetry, a nontrivial challenge would be to satisfy the consistency conditions\cite{CPt}. Now real $\mu\tau$ interchange antisymmetry\cite{Grimus} has been shown to arise in a class of explicit models with larger discrete symmetries including $\mathbb{Z}_4$ while Ref.\cite{joshi} discusses that the neutrino (Majorana) mass matrix can enjoy pure flavor antisymmetry under some discrete subgroups contained in $A_5$. Again, a real mixed $\mu\tau$ symmetry\cite{Joshipura:2015dsa} arises in a model where the charged lepton and neutrino mass matrices are invariant under specific residual symmetries contained in the finite discrete subgroups of $O(3)$. The latter work provides an explicit model based on $A_5$ maintaining the mixed $\mu\tau$ symmetry. However, such a demonstration is lacking in the literature for the corresponding CP-transformed (complex extended) cases.

The rest of the paper is organized as follows. Sec.\ref{sec2} deals with the symmetries of the neutrino Majorana mass matrix $M_\nu$ and the most general parametrization of $M_\nu$ that is invariant under \eqref{alum}. Sec.\ref{sec3} contains the evaluation of Majorana phases and a definite relation between the leptonic Dirac CP phase and the atmospheric mixing angle $\theta_{23}$ that involves the $\mu\tau$ mixing parameter $\theta$. In Sec.\ref{sec4} a numerical analysis of our proposal is presented utilizing neutrino oscillation data; this entails the extraction of the allowed parameter space and the prediction of light neutrino masses. It consists of three subsections. The first considers neutrinoless double beta decay; the second includes the range of variation of the CP asymmetry parameter $A_{\mu e}$ in experiments such as T2K, No$\nu$A and DUNE for both types of mass ordering; the variation of flavor flux ratios at neutrino telescopes is considered in the third. In Sec.\ref{sec5} we summarize the results of our analysis.

\section{Complex mixed $\mu\tau$ antisymmetry of the neutrino Majorana mass matrix}\label{sec2} The effective neutrino Majorana mass term in the Lagrangian density reads
\begin{equation} -\mathcal{L}_{mass}^\nu= \frac{1}{2}\overline{\nu_{Ll}^C} (M_\nu)_{lm}\nu_{Lm} + h.c. \label{lag}
\end{equation}  with $\nu_{Ll}^C=C\overline{(\nu_{Ll})}^T$ and the subscripts $l,m$ spanning the lepton flavor indices $e$, $\mu$, $\tau$ while the subscript $L$ denotes left-chiral neutrino fields. Here, $M_\nu$ is a complex symmetric matrix ($M_\nu^*\neq M_{\nu}=M_\nu^T$) in lepton flavor space. It can be diagonalized by a similarity transformation with a unitary matrix $U$:
\begin{equation} U^T M_\nu U=M_\nu^d \equiv {\rm diag}\hspace{1mm}(m_1,m_2,m_3).\label{e0} \end{equation} Here ${m}_i\hspace{1mm}(i=1,2,3)$ are real and we assume that $m_i\geq 0$. Without any loss of generality, we work in the diagonal basis of the charged leptons so that $U$ can be related to the PMNS matrix $U_{PMNS}$:
\bea
U=P_\phi U_{PMNS}\equiv
P_\phi \begin{pmatrix}
c_{1 2}c_{1 3} & e^{i\frac{\alpha}{2}} s_{1 2}c_{1 3} & s_{1 3}e^{-i(\delta - \frac{\beta}{2})}\\
-s_{1 2}c_{2 3}-c_{1 2}s_{2 3}s_{1 3} e^{i\delta }& e^{i\frac{\alpha}{2}} (c_{1 2}c_{2 3}-s_{1 2}s_{1 3} s_{2 3} e^{i\delta}) & c_{1 3}s_{2 3}e^{i\frac{\beta}{2}} \\
s_{1 2}s_{2 3}-c_{1 2}s_{1 3}c_{2 3}e^{i\delta} & e^{i\frac{\alpha}{2}} (-c_{1 2}s_{2 3}-s_{1 2}s_{1 3}c_{2 3}e^{i\delta}) & c_{1 3}c_{2 3}e^{i\frac{\beta}{2}}
\end{pmatrix},~\label{eu}
\eea
where $P_\phi={\rm diag}~(e^{i\phi_1},~e^{i\phi_2}~e^{i\phi_3})$ is an unphysical diagonal  phase matrix and  $c_{ij}\equiv\cos\theta_{ij}$, $s_{ij}\equiv\sin\theta_{ij}$ with the mixing angles $\theta_{ij}\in[0,\pi/2]$. We follow the PDG convention\cite{Tanabashi:2018oca} but denote our Majorana phases by $\alpha$ and $\beta$ instead of $\alpha_{21}$ and $\alpha_{31}$. CP-violation enters through nontrivial values of the Dirac phase $\delta$ and of the Majorana phases $\alpha,\beta$  with $\delta,\alpha,\beta\in[0,2\pi]$.\\

The effect of our proposed invariance under \eqref{alum} on the neutrino Majorana mass matrix would be \begin{equation} \mathcal{G}^{\theta T}M_\nu \mathcal{G}^\theta=-M_\nu^*\label{aff}.\end{equation} $\mathcal{G}^\theta$ in \eqref{aff} is given by $iG^\theta$ where $G^\theta$ was defined in \eqref{chat}. In flavor space, the most generally parameterized $3\times 3$ complex symmetric mass matrix obeying \eqref{aff} is given by \begin{equation}M^{CP\theta A}_\nu=\begin{pmatrix} ix & a_1+ia_2 & a_1t^{-1}_{\frac{\theta}{2}}-ia_2t_{\frac{\theta}{2}} \\a_1+ia_2 & y_1+iy_2 & y_1c_\theta s_\theta^{-1}+ic\\ a_1t^{-1}_{\frac{\theta}{2}}-ia_2t_{\frac{\theta}{2}} & y_1c_\theta s_\theta^{-1}+ic & -y_1+i(y_2+2cc_\theta s_\theta^{-1})
\end{pmatrix},\label{asp}\end{equation} where $c_\theta\equiv\cos\theta, s_\theta\equiv \sin\theta$ and $t_{\frac{\theta}{2}}\equiv\tan\frac{\theta}{2}$. In \eqref{asp}, there are seven real free parameters $x,a_{1,2},c,y_1,y_2$ and $\theta$. As expected, the limit $\theta\to\pi/2$ gives back the mass matrix $M^{CP\mu\tau A}_{\nu}$ invariant under CP transformed $\mu\tau$ interchange antisymmetry\cite{Samanta:2017kce}, namely\begin{equation}
M_{\nu}^{CP\mu\tau A}=\begin{pmatrix} ix & a_1+ia_2 & a_1-ia_2\\a_1+ia_2 & y_1+iy_2 & ic\\ a_1-ia_2 & ic & -y_1+iy_2\end{pmatrix}.\end{equation}

It should be emphasized that complex mixed $\mu\tau$ antisymmetry, which can be abbreviated as $CP^{\theta\mu\tau A}$ and gets generated by $\mathcal{G}^\theta$, must now be broken in the charged lepton sector. This is because a nonzero Dirac CP violation is equivalent to the criterion\begin{equation}{\rm Tr}~[H_\nu,H_\ell]^3\neq 0, \label{r4}
\end{equation} where $H_\nu$ and $H_\ell$ are two hermitian matrices defined as $H_\ell=M_\ell^\dagger M_\ell$, $M_\ell$ being the charged lepton mass matrix and $H_\nu=M_\nu^\dag M_\nu$. \cite{Bernabeu:1986fc}. A common CP symmetry $\mathcal{G}_{CP}$ would imply \begin{equation}\mathcal{G}_{CP}^T H_\nu^T \mathcal{G}_{CP}^*= H_\nu,~~\mathcal{G}_{CP}^T H_\ell^T \mathcal{G}_{CP}^*= H_\ell. \label{r5}\end{equation}From \eqref{r5} it follows that  ${\rm Tr}[H_\nu,H_\ell]^3= 0$  which leads to $\sin \delta =0$ i.e. a vanishing Dirac CP violation. As mentioned earlier, this is disfavored by current experiments.

\section{Neutrino mixing angles and phases}\label{sec3} Eqs.\eqref{e0} and \eqref{aff} together imply\cite{Grimus:2003yn} that \begin{equation}\mathcal{G}^\theta U^*=U\tilde{d}\label{atto}\end{equation} where $\tilde{d}_{ij}=\pm \delta_{ij}.$ Next, we take $\tilde{d}={\rm diag}(\tilde{d}_1,\tilde{d}_2,\tilde{d}_3)$ where each $\tilde{d}_i$ $(i = 1, 2, 3)$ is either $+1$ or $-1$. \eqref{atto} can explicitly be written as
\begin{equation}\begin{pmatrix}-i & 0 & 0\\0 & -ic_\theta & is_\theta\\0 & is_\theta & ic_\theta \end{pmatrix}\begin{pmatrix}
U^*_{e1} & U^*_{e2} & U^*_{e 3}\\U^*_{\mu 1} & U^*_{\mu2} & U^*_{\mu3}\\
U^*_{\tau1} & U^*_{\tau2} & U^*_{\tau3}
\end{pmatrix}=\begin{pmatrix}
\tilde{d}_1 U_{e1} & \tilde{d}_2 U_{e2} & \tilde{d}_3 U_{e 3}\\
\tilde{d}_1 U_{\mu 1} & \tilde{d}_2 U_{\mu2} & \tilde{d}_3 U_{\mu3}\\
\tilde{d}_1 U_{\tau1} & \tilde{d}_1 U_{\tau2} & \tilde{d}_1 U_{\tau3}
\end{pmatrix}.\label{oju}
\end{equation} Eq. \eqref{oju} leads to nine independent relations corresponding to the three rows: $$
-iU^*_{e1}=\tilde{d}_1U_{e1},~-iU^*_{e2}=\tilde{d}_2U_{e2},~-iU^*_{e2}=\tilde{d}_3U_{e3},$$ $$-iU^*_{\mu 1}c_\theta+iU^*_{\tau 1}s_\theta=\tilde{d}_1U_{\mu 1},~-iU^*_{\mu 2}c_\theta+iU^*_{\tau 2}s_\theta=\tilde{d}_2U_{\mu 2},~ -iU^*_{\mu 3}c_\theta+iU^*_{\tau 3}s_\theta=\tilde{d}_3U_{\mu 3},$$
\begin{equation}
iU^*_{\mu 1}s_\theta+iU^*_{\tau 1}c_\theta=\tilde{d}_1U_{\tau 1},~iU^*_{\mu 2}s_\theta+iU^*_{\tau 2}c_\theta=\tilde{d}_2U_{\tau 2},~iU^*_{\mu 3}s_\theta+iU^*_{\tau 3}c_\theta=\tilde{d}_3U_{\tau 3}.\label{relo}
\end{equation} In order to calculate the Majorana phases in a way that avoids unphysical phases, it is useful to construct two rephasing invariants\cite{Branco} \begin{equation}
I_1=U_{e1}U^*_{e2}, I_2=U_{e1}U^*_{e3}.\label{3}
\end{equation} Using the relations in the first row of \eqref{relo}, we obtain \begin{equation}
I_1=\tilde{d}_1\tilde{d}_2U^*_{e1}U_{e2},~ I_2=\tilde{d}_1\tilde{d}_2U^*_{e1}U_{e3}.\label{4}\end{equation} On inserting the two different expressions for $I_{1,2}$, in \eqref{3} and \eqref{4}, we find that \begin{equation}
c_{12}s_{12}c^2_{13}e^{-i\alpha/2}=\tilde{d}_1\tilde{d}_2 c_{12}s_{12}c^2_{13}e^{i\alpha/2}\label{5}
\end{equation} and \begin{equation}
c_{12}s_{13}c_{13}e^{i(\delta-\beta/2)}=\tilde{d}_1\tilde{d}_3 c_{12}s_{13}c_{13}e^{-i(\delta-\beta/2)}\label{6}.
\end{equation} From \eqref{5} and \eqref{6}, it follows that \begin{equation}
e^{i\alpha}=\tilde{d}_1\tilde{d}_2,~ e^{2i(\delta-\beta/2)}=\tilde{d}_1\tilde{d}_3,
\end{equation} i.e., either $\alpha=0$ or $\alpha=\pi$, and either $\beta=2\delta$ or $\beta=2\delta-\pi$. In other words, the Majorana phases can have four possible pairs of values for a given value of $\delta$. From the absolute square of the third relation in the third row of \eqref{relo}, we obtain \begin{equation}
|U_{\tau 3}|^2=(U^*_{\mu 3}s_\theta+U^*_{\tau 3}c_\theta)(U_{\mu 3}s_\theta+U_{\tau 3}c_\theta)\end{equation} which implies that\begin{equation}
\cot2\theta_{23}=\cot\theta\cos(\phi_2-\phi_3)
\end{equation} reducing to $\theta_{23}\to\pi/4$ in the $\mu\tau$ interchange limit $\theta\to\pi/2$, as expected. Taking the absolute square of the second relation in the third row of \eqref{relo}, and eliminating the unphysical phase difference $\phi_2-\phi_3$, we obtain\begin{equation}
\sin\delta=\pm\sin\theta/\sin2\theta_{23}.\label{xop}
\end{equation} This result was originally derived in Ref.\cite{Chen:2015siy} which proposed a CP transformed mixed $\mu\tau$ symmetry for neutrinos. Eq.\eqref{xop}, as expected, reproduces the result $\sin\delta=\pm 1$ (equivalently, $\cos\delta=0$) in the $\mu\tau$ interchange limit $\theta=\pi/2$ and $\theta_{23}=\pi/4$. Note also that, if the unphysical phase combination $\phi_2-\phi_3$ is put equal to zero, $\cot2\theta_{23}$ becomes equal to $\cot\theta$ and $\cos\delta$ vanishes i.e., leptonic Dirac CP violation becomes maximal. However, such is not the case in general. We should also mention that another relation between $\delta$ and $\theta_{13}$ was obtained recently in Ref.\cite{Rahat:2018sgs}.

\section{Numerical analysis}\label{sec4} 
\begin{figure}[H]
\begin{center}
\includegraphics[scale=.20]{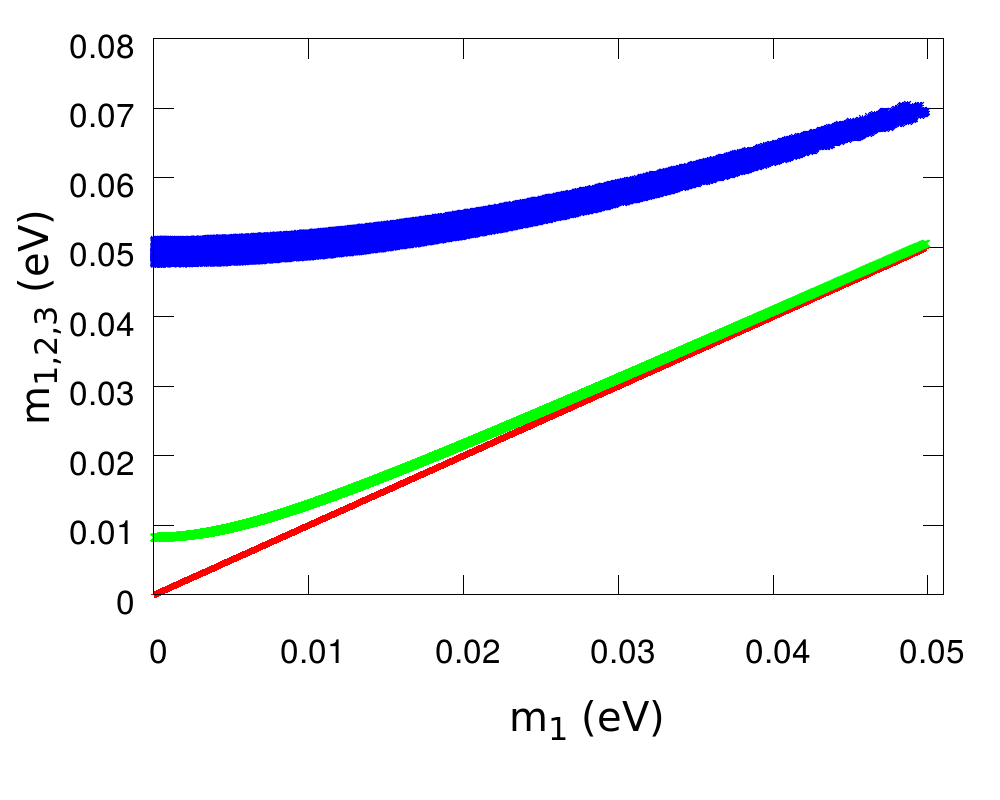}\includegraphics[scale=.20]{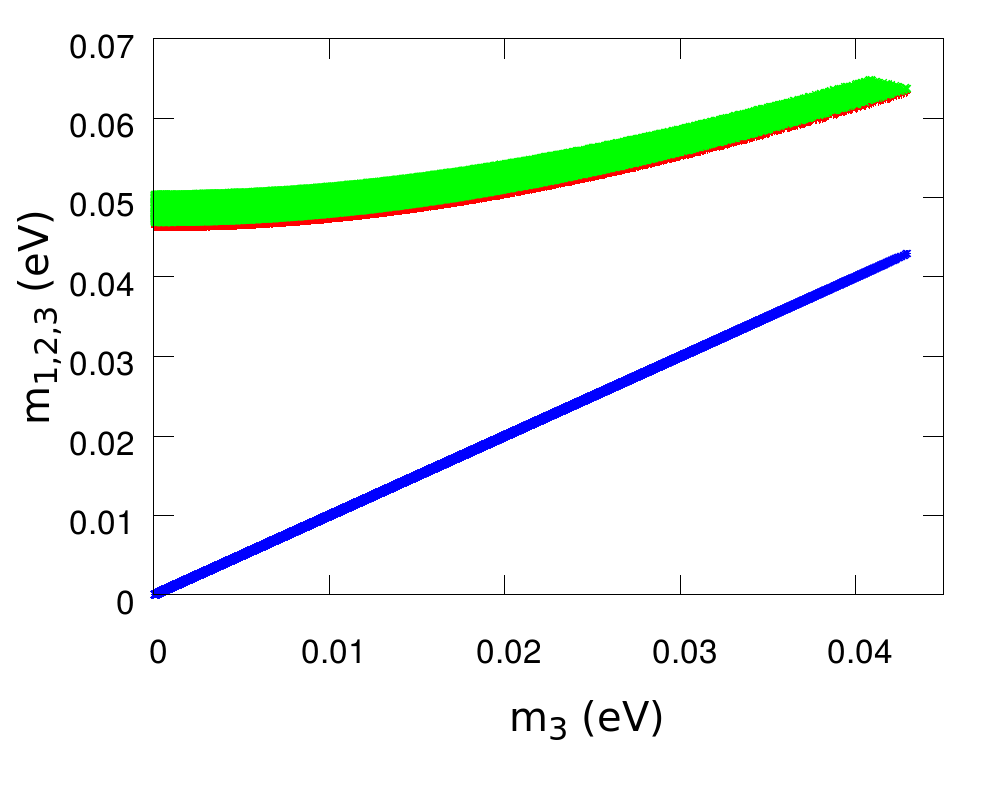}
\caption{Plots of $m_{1,2,3}$ for normal (left) and inverted (right) mass ordering with the lightest mass eigenvalue is plotted in the ordinate. The red, green and blue bands refer to $m_1,m_2$ and $m_3$ respectively.}\label{fg6}
\end{center}
\end{figure}

In order to demonstrate the phenomenological viability of our theoretical proposal we present a numerical analysis of its consequences in substantial detail. It is organized as follows. In Table \ref{oscx}, we display  the $3\sigma$ ranges of neutrino mixing angles and mass squared differences obtained from globally fitted neutrino oscillation data\cite{Esteban:2016qun}. The allowed ranges of parameters of $M_\nu$, CP phases and the consequent predictions on the light neutrino masses are tabulated in Table \ref{osc2}, \ref{oscnew} and Table \ref{t3} respectively. These have been obtained by using the exact analytical formulae for the mixing angles and light neutrino masses\cite{Adhikary:2013bma}, the entries in Table \ref{oscx} and the upper bound\cite{Aghanim:2016yuo} of $0.17$ eV on the sum of the light neutrino masses from PLANCK and other cosmological observations. In Fig.\ref{fg6} each mass eigenvalue $m_{1}, m_2$ and $m_3$ is plotted against the smallest mass eigenvalue $m_{min}$ for both types of mass ordering. The neutrino mass spectrum is clearly hierarchical ($m_{2,1}\gg m_3$ for NO and $m_{2,1}\ll m_3$ for IO).
\begin{table}[H]
\begin{center}
\caption{Input values used in the analysis\cite{Esteban:2016qun}} \label{oscx}
 \begin{tabular}{|c|c|c|c|c|c|}
\hline
${\rm Parameter}$&$\theta_{12}$&$\theta_{23}$ &$\theta_{13}$ &$ \Delta m_{21}^2$&$|\Delta m_{31}^2|$\\
&$\rm degrees$&$\rm degrees$ &$\rm degrees$ &$ 10^{-5}\rm (eV)^2$&$10^{-3} \rm (eV^2)$\\
\hline
$3\sigma\hspace{1mm}{\rm ranges\hspace{1mm}(NO)\hspace{1mm}}$&$31.42-36.05$&$40.3-51.5$&$8.09-8.98$&
$6.80-8.02$&$2.399-2.593$\\
\hline
$3\sigma\hspace{1mm}{\rm ranges\hspace{1mm}(IO)\hspace{1mm}}$&$31.43-36.06$&$41.3-51.7$&$8.14-9.01$&
$6.80-8.02$&$2.369-2.562$\\
\hline
${\rm Best\hspace{1mm}{\rm fit\hspace{1mm}}values\hspace{1mm}(NO)}$ & $33.62$ & $47.2$ &  $8.54$ &$7.40$ & $2.494$\\
\hline
${\rm Best\hspace{1mm}{\rm fit\hspace{1mm}}values\hspace{1mm}(IO)}$&$33.62$&$48.1$&$8.58$&$7.40$&$2.465$\\
\hline
\end{tabular}
\end{center}
\end{table}
\noindent

\begin{table}[H]
\begin{center}
\caption{Output values of the parameters of $M_\nu$} \label{osc2}
 \begin{tabular}{|c|c|c|c|c|c|c|c|}
\hline
${\rm Parameters}$& $x/10^{-2}$&$a_1/10^{-2}$ &$a_2/10^{-2}$ &$y_1/10^{-2}$&$y_2/10^{-2}$&$c/10^{-2}$ & $\theta (^\circ)$\\
\hline
${\rm NO}$& -$2.2$ $-$ $2.2$&-$4.5$ $-$ $4.5$&-$3.2$ $-$ $3.2$&-$3.5$ $-$ $3.5$&-$4.5$ $-$ $4.5$
&-$3.5$ $-$ $3.5$ & $12$-$174$\\
\hline
${\rm IO}$&-$2.5$ $-$ $2.5$&-$4.5$ $-$ $4.5$&-$0.4$ $-$ $0.4$&-$2.5$ $-$ $2.5$&-$3.5$ $-$ $3.5$
&-$2.5$ $-$ $2.5$& $2$-$156$\\
\hline
\end{tabular}
\end{center}
\end{table}
\noindent

\begin{table}[H]
\begin{center}
\caption{Output values CP phases in the range $\beta\in[0,2\pi]$)} \label{oscnew}
 \begin{tabular}{|c|c|c|c|c|c|c|c|}
\hline
${\rm Ordering}$& $\delta$& $\beta=2\delta$ &$\beta=2\delta-\pi$\\
\hline
${\rm NO} (\sin\delta>0)$& $[6^\circ,174^\circ]$ & $[12^\circ,348^\circ]$  & $[0^\circ,168^\circ],[192^\circ,360^\circ]$\\
\hline
${\rm NO} (\sin\delta<0)$ & $[186^\circ,354^\circ]$ & $[12^\circ,348^\circ]$ & $[0^\circ,168^\circ],[192^\circ,360^\circ]$\\
\hline
$ {\rm IO} (\sin\delta>0)$& $[4^\circ,176^\circ]$ & $[8^\circ,352^\circ]$  & $[0^\circ,172^\circ],[188^\circ,360^\circ]$\\
\hline
${\rm IO} (\sin\delta<0)$ & $[184^\circ,356^\circ]$ & $[8^\circ,352^\circ]$ & $[0^\circ,172^\circ],[188^\circ,360^\circ]$\\
\hline
\end{tabular}
\end{center}
\end{table}
\noindent

\begin{table}[H]
\begin{center}
\caption{Predictions on the light neutrino masses.} \label{t3}
 \begin{tabular}{|c|c|c|c|c|c|}
 \hline
  \multicolumn{3}{|c|}{{Normal Ordering} $(m_3>m_2)$} & \multicolumn{3}{c|}{{Inverted Ordering $(m_3<m_1)$}} \\
\hline
$m_1/10^{-3}$&$m_2/10^{-3}$ &$m_3/10^{-3}$ &$ m_1/10^{-3}$&$m_2/10^{-3}$&$m_3/10^{-3}$ \\
$\rm (eV)$&$ \rm (eV)$ &$\rm (eV)$ &$\rm (eV)$&$ \rm (eV)$&$\rm (eV)$\\
\hline
$8.4\times 10^{-2}-49$&$9-51$&$50-71$&$48-64$&$49-66$&$4.4\times 10^{-2}-42$\\
\hline
\end{tabular}
\end{center}
\end{table}
\noindent

Next, we discuss the numerical results of CP-transformed mixed $\mu\tau$ antisymmetry for neutrinoless double beta decay, CP asymmetry in neutrino oscillations and flavor flux ratios at neutrino telescopes in three separate subsections.

\textbf{Neutrinoless double beta decay}- In this subsection, we explore the predictions of our proposal for $\beta\beta 0\nu$ decay. The latter is a lepton number violating process arising from the decay of a nucleus as\begin{equation}(A,Z)\longrightarrow (A, Z+2)+2e^-\end{equation} characterized by the absence of any final state neutrinos. The observation of such a decay will lead to the confirmation of the Majorana nature of the neutrinos. The half-life\cite{Rodejohann:2011mu} corresponding to the above decay is given by
\begin{equation}
\frac{1}{T^{0\nu}_{1/2}}=G_{0\nu}|\mathcal{M}|^2 |M_\nu^{ee}|^2m_e^{-2},
\end{equation}
where $G_{0\nu}$ is the two-body phase space factor, $\mathcal{M}$ is the nuclear matrix element (NME), $m_e$ is the mass of the electron and  $M_\nu^{ee}$ is the (1,1) element of the effective light neutrino mass matrix $M_\nu$. In the PDG parametrization convention for $U_{\rm PMNS}$, $M^{ee}_\nu$ is given most generally by
\begin{equation}
M^{ee}_\nu=c_{12}^2c_{13}^2m_1+s_{12}^2c_{13}^2m_2e^{i\alpha}+
s_{13}^2m_3e^{i(\beta-2\delta)}.\label{alps}
\end{equation}
\noindent

\begin{figure}[H]
\includegraphics[scale=.25]{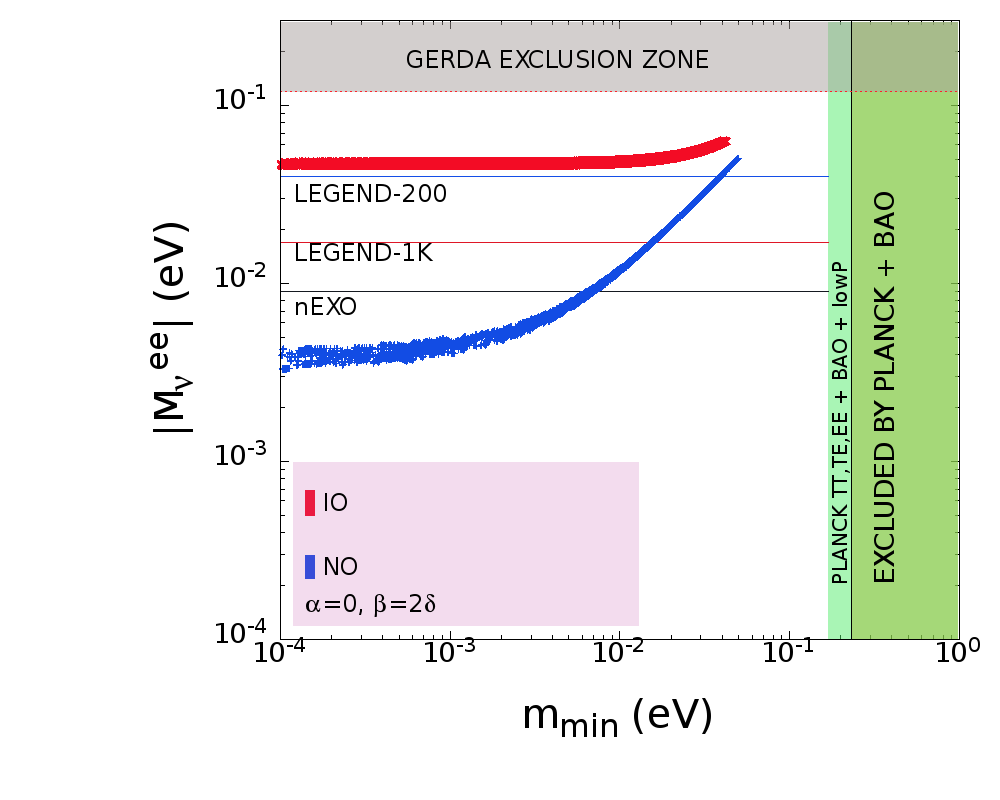}\includegraphics[scale=.25]{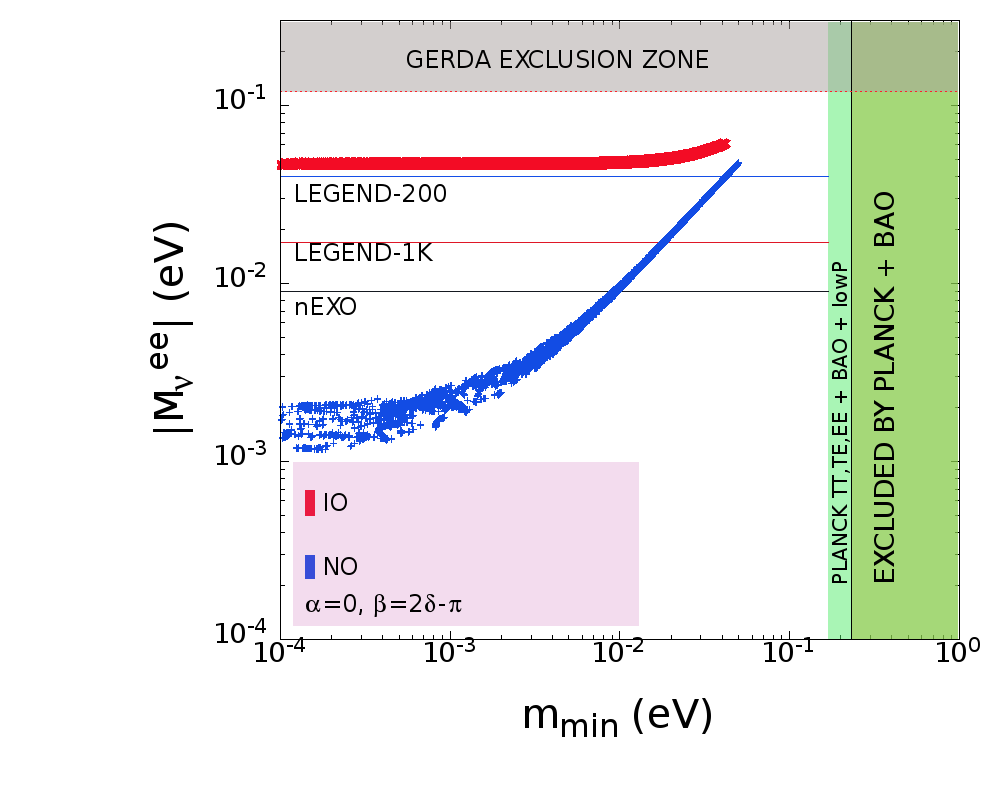}\\
\includegraphics[scale=.25]{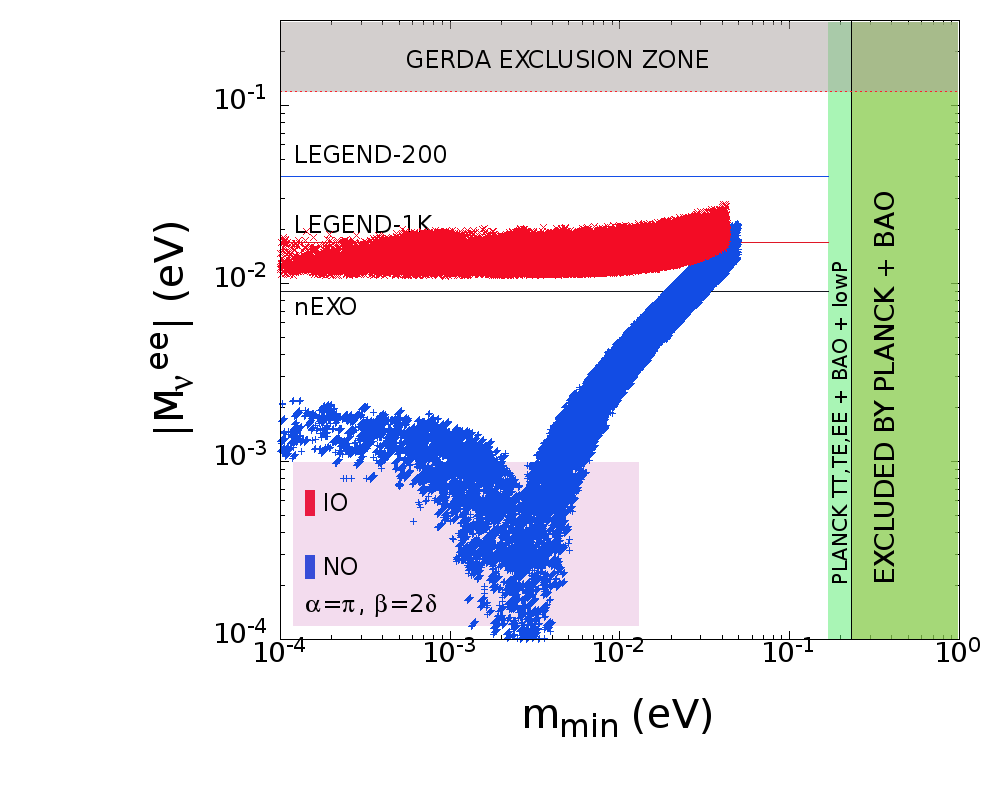}\includegraphics[scale=.25]{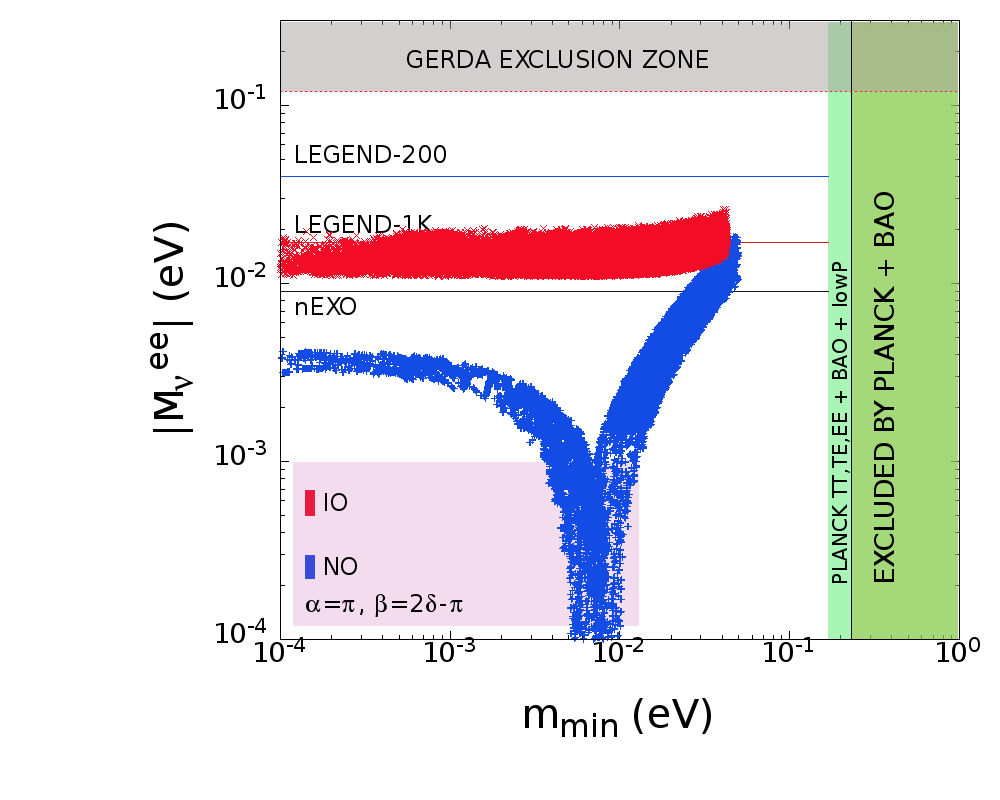}
\caption{Plots of $|M^{ee}_\nu|$ vs. $m_{min}$ for both types of mass ordering with four possible choices of the Majorana phases $\alpha$ and $\beta$. NO and IO refer to Normal and Inverted ordering respectively.}\label{fg5}
\end{figure}
\noindent

In our case, \eqref{alps} simplifies to the following four expressions for our four different possibilities:
\vspace{0.5cm}

(i) $|M^{ee}_\nu|=c^2_{12}c^2_{13}m_1+s^2_{12}c^2_{13}m_2+s^2_{13}m_3$ for $\alpha=0,\beta=2\delta$,

(ii)$|M^{ee}_\nu|=c^2_{12}c^2_{13}m_1+s^2_{12}c^2_{13}m_2-s^2_{13}m_3$ for  $\alpha=0,\beta=2\delta-\pi$,

(iii) $|M^{ee}_\nu|=c^2_{12}c^2_{13}m_1-s^2_{12}c^2_{13}m_2+s^2_{13}m_3$ for $\alpha=\pi,\beta=2\delta$

and

(iv) $|M^{ee}_\nu|=c^2_{12}c^2_{13}m_1-s^2_{12}c^2_{13}m_2-s^2_{13}m_3$ for $\alpha=\pi,\beta=2\delta-\pi$.
\noindent

In $0\nu\beta\beta$ decay, $M^{ee}_\nu$ depends on $\alpha$ and $\beta-2\delta$ (cf. Eq.(4.3)). In a generic case, $\alpha$ and $\beta-2\delta$ varies in the range $[0,2\pi]$ (or $[-\pi,\pi]$, since angles are defined modulo $2\pi$) to cover the largest possible parameter space. However, a notable feature of our scenario is that it uniquely fixes (i) $\alpha$ to be $0$ or $\pi$ and (ii) the combination $\beta-2\delta$ to $0$ or $-\pi$ rather than the entire range of variation $[0,2\pi]$ (or $[-\pi,\pi]$) as in a generic situation. This constraint tightly controls the range of variation of $M_\nu^{ee}$ and is implicitly reflected in the parameter space of $0\nu\beta\beta$ decay. The resulting plots of $|M^{ee}_\nu|$ versus the smallest mass eigenvalue $m_{min}$ ($m_1$ for NO and $m_3$ for IO) are presented in Fig.\ref{fg5} with significant upper limits on $|M^{ee}_\nu|$ for ongoing and future experiments. At the moment the most stringent exclusion zone on $M_{ee}$ has been reported by the GERDA Phase II\cite{gerda2} experiment to be $0.12-0.26$eV depending on the value of the nuclear matrix element used. It is evident from Fig.\ref{fg5} that  $|M_{ee}|$ in each plot leads to an upper limit which is below the reach of the GERDA phase-II experimental data. The sensitivity reach of several other experiments such as LEGEND-200 (40 meV), LEGEND-1K (17 meV) and nEXO (9 meV)\cite{Agostini:2017jim}, shown in Fig.\ref{fg5}, can probe our model. In particular, if LEGEND-1K fails to observe a signal, the inverted mass ordering in our model corresponding to $\alpha=0$ shall be excluded. Note that, for each case, the entire parameter space corresponding to the inverted mass ordering is likely to be ruled out for both $\alpha=0$ and $\pi$ if nEXO, covering its entire reach, does not observe any $\beta\beta0\nu$ signal. However, the latter exclusion is likely to be a generic feature of many models.

\textbf{CP asymmetry in neutrino oscillations}- Here we discuss the effect of the existence of leptonic Dirac CP violation $\delta$ in neutrino oscillation experiments. The phase $\delta$ makes its appearance in the CP asymmetry parameter $A_{lm}$, defined as \begin{equation}
A_{lm}=\frac{P(\nu_l\to\nu_m)-P(\bar{\nu}_l\to\bar{\nu}_m)}{P(\nu_l\to\nu_m)+P(\bar{\nu}_l\to\bar{\nu}_m)},
\end{equation} where $l,m=(e,\mu,\tau)$ are flavor indices and the $P$'s are transition probabilities. The $\nu_\mu\to \nu_e$ transition probability is given by \begin{equation}P_{\mu e}\equiv P(\nu_\mu\to\nu_e)=P_{atm}+P_{sol}+2\sqrt{P_{atm}}\sqrt{P_{sol}}\cos(\Delta_{32}+\delta).\label{x}\end{equation} where $\Delta_{ij}=\Delta m^2_{ij}L/4E$ is the kinematic phase factor in which $L$ denotes the baseline length and $E$ represents the beam energy. The quantities $P_{atm},P_{sol}$ are respectively defined as \begin{eqnarray}
\sqrt{P_{atm}}=\sin\theta_{23}\sin2\theta_{13}\frac{\sin(\Delta_{31}-aL)}{(\Delta_{31}-aL)}\Delta_{31},\\
\sqrt{P_{sol}}=\cos\theta_{23}\sin2\theta_{12}\frac{\sin aL}{aL}\Delta_{21},\end{eqnarray} where $a=G_F N_e/\sqrt{2}$ with $G_F$ being the Fermi constant and $N_e$ being the electron number density in the medium of propagation which takes into account the matter effects in neutrino propagation through the earth.  An approximate value of $a$ for the earth is $3500{\rm km}^{-1}$. In the limit $a\to 0$, \eqref{x} leads to the oscillation probability in vacuum. With this, the CP asymmetry parameter is given by \begin{equation}
A_{\mu e}=\frac{P(\nu_\mu\to\nu_e)-P(\bar{\nu}_\mu\to\bar{\nu}_e)}{P(\nu_\mu\to\nu_e)+P(\bar{\nu}_\mu\to\bar{\nu}_e)}=
\frac{2\sqrt{P_{atm}}\sqrt{P_{sol}}\sin\Delta_{32}\sin\delta}{P_{atm}+2\sqrt{P_{atm}}\sqrt{P_{sol}}\cos\Delta_{32}\cos\delta+P_{sol}}
\end{equation} where $\sin\delta$, given by \eqref{xop}, has two possible values and same goes for $\cos\delta$. Hence there are two pairs of choices which give rise to two pairs of possibilities for $A_{\mu e}$ as given in Table \ref{axe}.

\begin{table}[H]
\begin{center}
\caption{Four possibilities for $A_{\mu e}$} \label{axe}
 \begin{tabular}{|c|c|c|}
\hline
${\rm Possibilities}$&$\sin\delta$&$\cos\delta$\\
\hline
${\rm Case~ A}$&$+\sin\theta(\sin2\theta_{23})^{-1}$&$+(\sin2\theta_{23})^{-1}\sqrt{\cos^2\theta\sin^22\theta_{23}-\sin^2\theta\cos^22\theta_{23}}$\\
\hline
${\rm Case~ B}$&$-\sin\theta(\sin2\theta_{23})^{-1}$&$+(\sin2\theta_{23})^{-1}\sqrt{\cos^2\theta\sin^22\theta_{23}-\sin^2\theta\cos^22\theta_{23}}$\\
\hline
${\rm Case~ C}$ & $+\sin\theta(\sin2\theta_{23})^{-1}$ & $-(\sin2\theta_{23})^{-1}\sqrt{\cos^2\theta\sin^22\theta_{23}-\sin^2\theta\cos^22\theta_{23}}$\\
\hline
${\rm Case~ D}$&$-\sin\theta(\sin2\theta_{23})^{-1}$&$-(\sin2\theta_{23})^{-1}\sqrt{\cos^2\theta\sin^22\theta_{23}-\sin^2\theta\cos^22\theta_{23}}$\\
\hline
\end{tabular}
\end{center}
\end{table}
\noindent

\begin{figure}[H]
\begin{center}
\includegraphics[scale=.20]{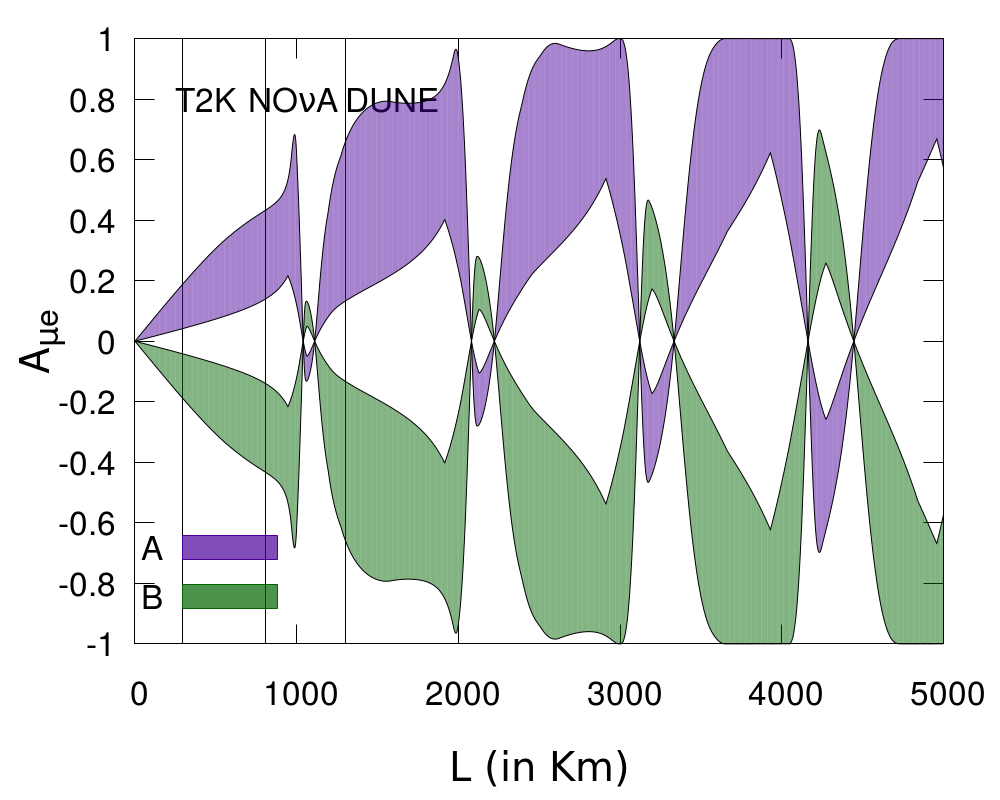}\includegraphics[scale=.20]{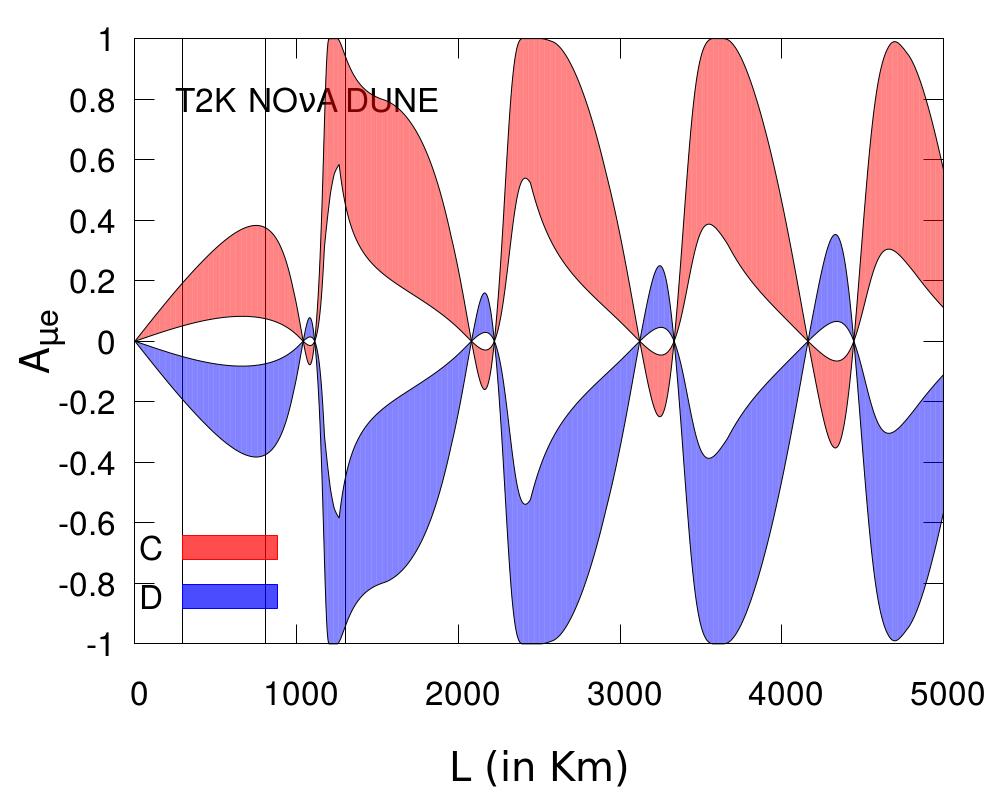}
\caption{CP asymmetry parameter $A_{\mu e}$ (for $E=1$ GeV), plotted against the baseline length $L$, for the four possibilities in Table \ref{axe}. Each plot stands for both NO and IO since numerically, within the $3\sigma$ range of $\theta_{23}$, the two types of ordering are practically indistinguishable. The bands are due to $\theta_{23}$ and $\theta$ being allowed to vary within their experimental $3\sigma$ range and phenomenologically allowed range respectively with the other parameters kept at their best fit values.}\label{fig8}
\end{center}
\end{figure}

In Fig.\ref{fig8} the CP asymmetry parameter $A_{\mu e}$, for both types of mass ordering, is plotted against the baseline length $L$ for four possibilities (Table \ref{axe}) and for a fixed beam energy ($E=1$ GeV). The baseline lengths corresponding to experiments such as T2K, No$\nu$A and DUNE have been shown by vertical lines in the figure. For concreteness, Table 6 provides the range of variation of CP asymmetry parameter $A_{\mu e}$ for a fixed energy of $E=1$GeV in T2K, NO$\nu$A and DUNE.

\begin{figure}[H]
\begin{center}
\includegraphics[scale=.20]{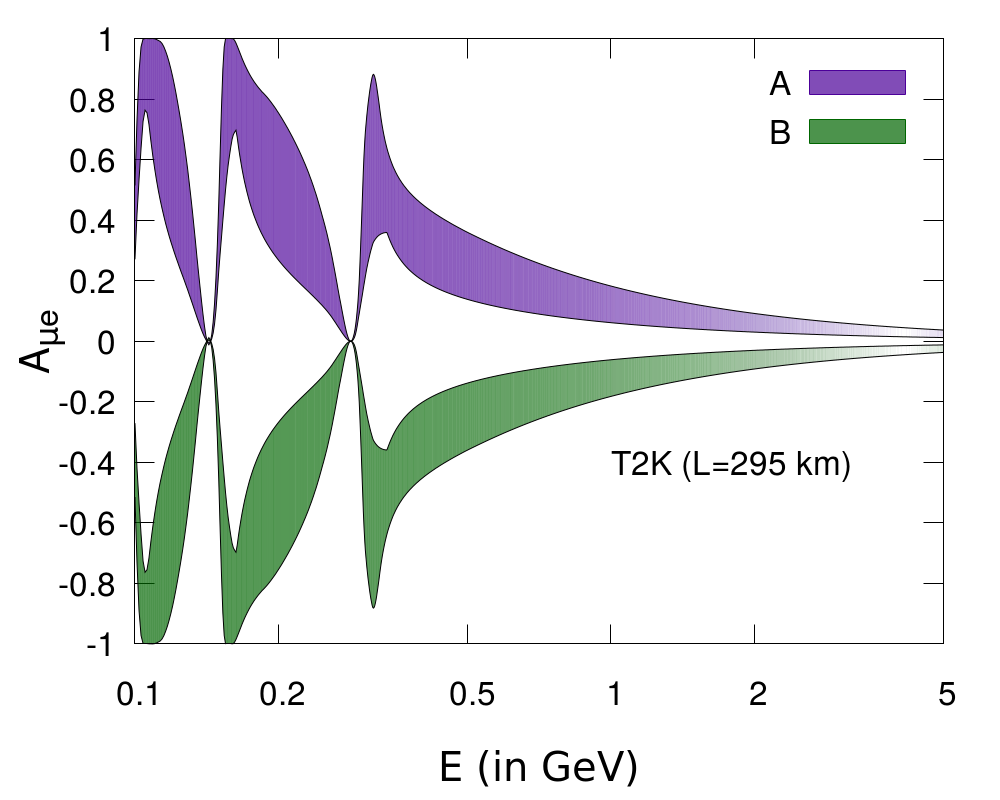}\includegraphics[scale=.20]{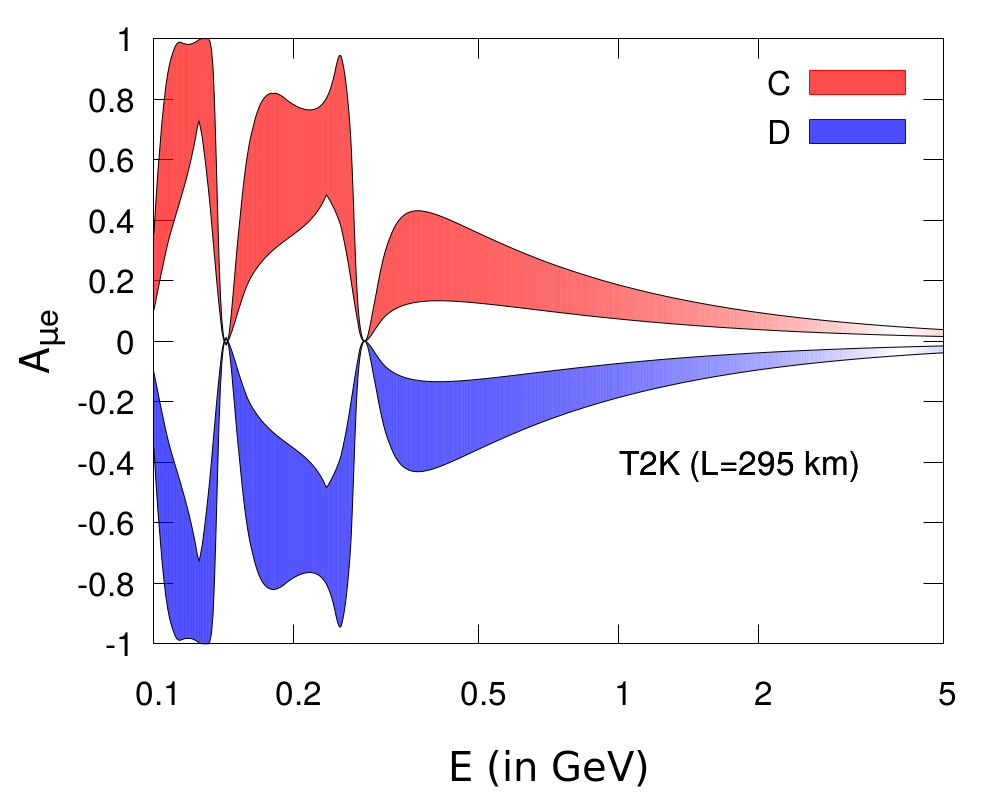}

\includegraphics[scale=.20]{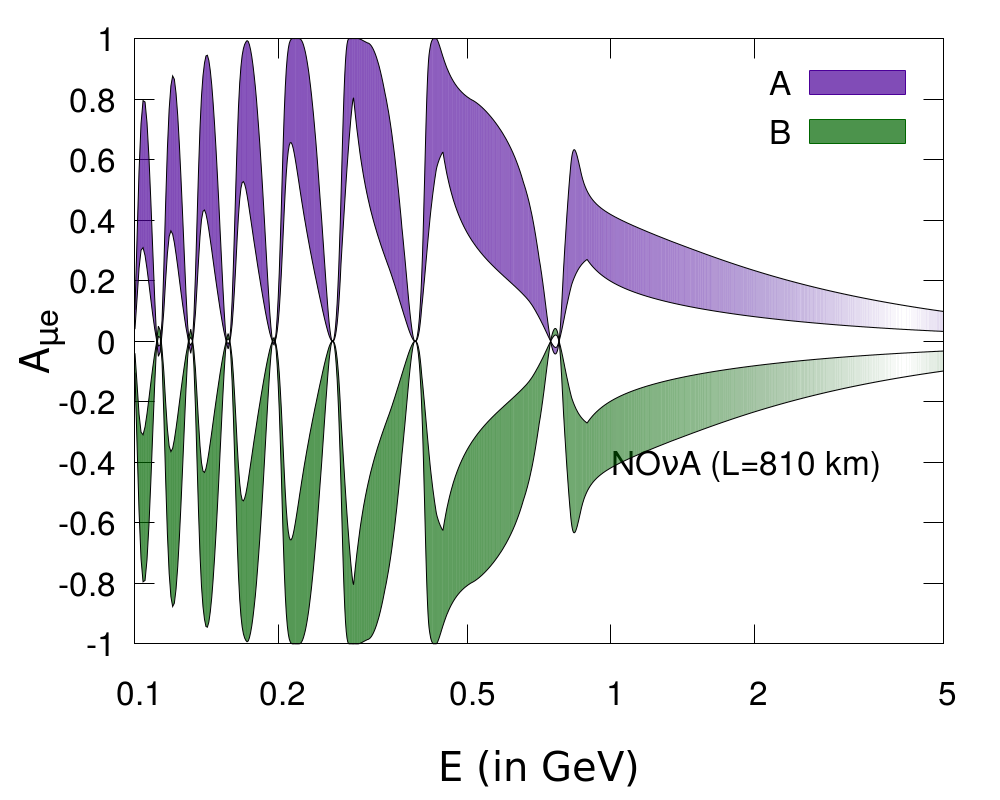} \includegraphics[scale=.20]{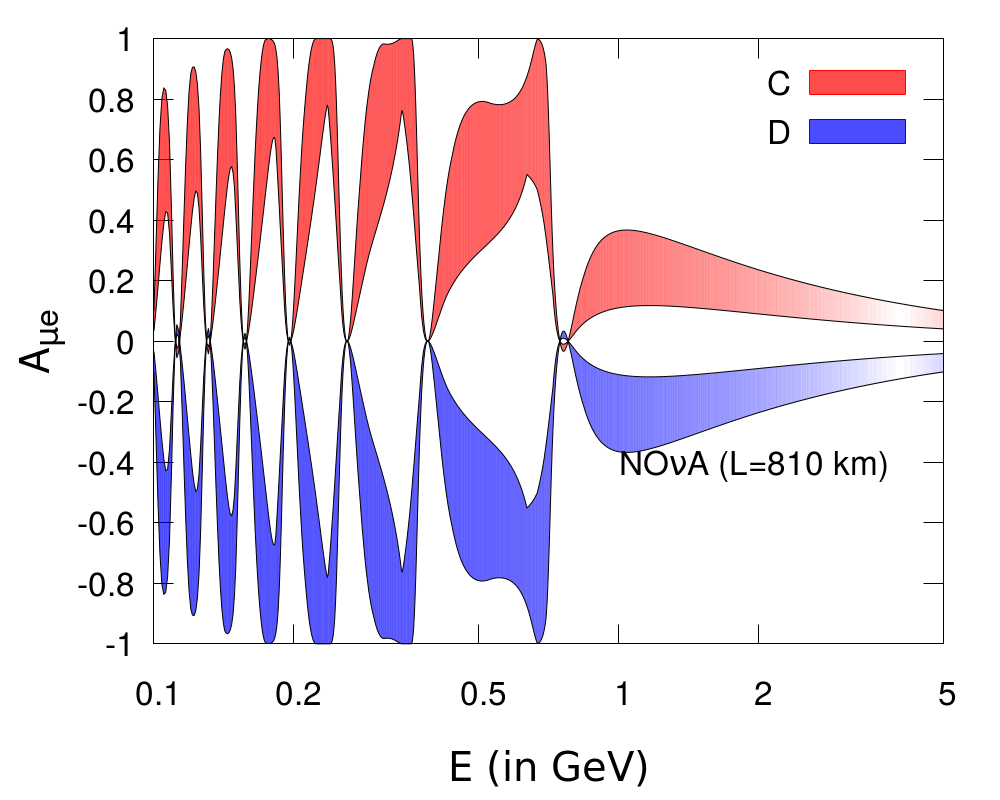}

\includegraphics[scale=.20]{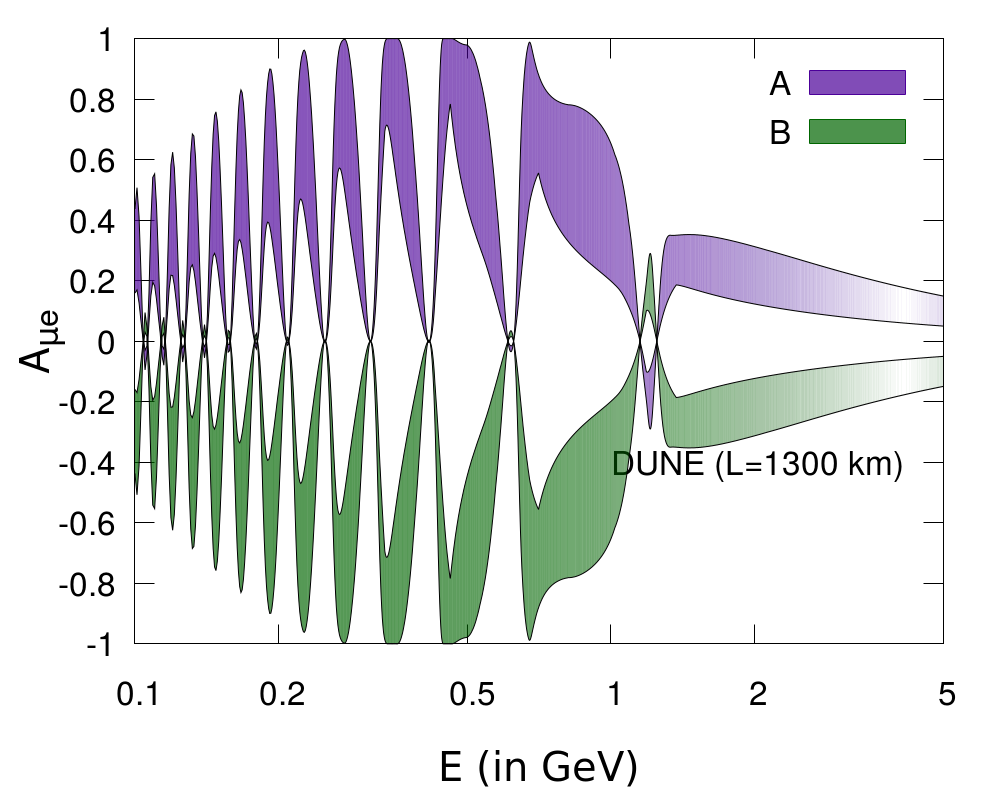}\includegraphics[scale=.20]{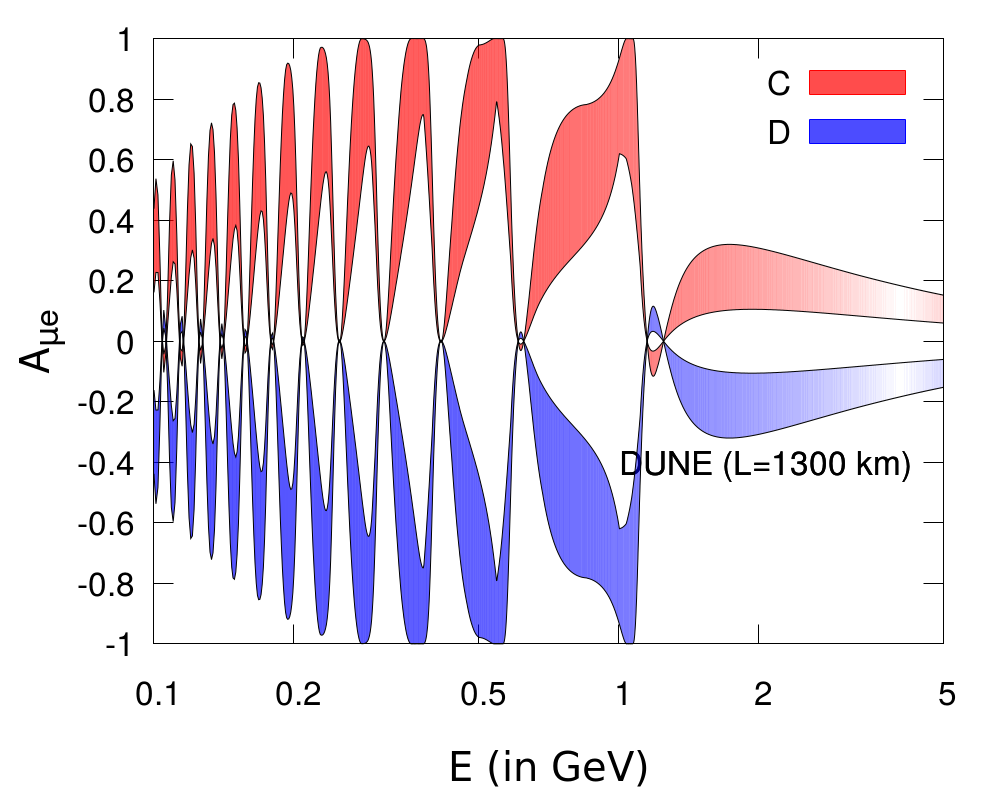}

\caption{Variation of the CP asymmetry parameter with beam energy $E$ for different baselines lengths of $L=295$ km, $810$ km and $1300$ km corresponding to T2K, NO$\nu$A and DUNE respectively for both NO and IO; the numerical distinction between the two types of ordering is insignificant for the $3\sigma$ range of $\theta_{23}$.}\label{fig9a}
\end{center}
\end{figure}

In Fig.\ref{fig9a}, $A_{\mu e}$ is plotted against the beam energy $E$ for four possible cases (Table \ref{axe}) separately for T2K, No$\nu$A and DUNE for both types of mass ordering. In generating these plots, the atmospheric mixing angle $\theta_{23}$ has been taken to be within its currently allowed $3\sigma$ range while the remaining neutrino oscillation parameters have  been kept fixed at their best fit values. For each of these experiments, Table 7 summarizes the allowed ranges of $A_{\mu e}$ for different values of the energy $E$.
%
%
%
%

\begin{table}[H]
\begin{center}
\caption{Prediction of the ranges of $|A_{\mu e}|$ with $E=1$GeV} \label{cpas1}
\begin{tabular}{|c|c|c|c|}
\hline
${\rm Experiment}$ & ${\rm T2K}$ & ${\rm NO}\nu{\rm A}$ & ${\rm DUNE}$\\
\hline
${\rm Case ~A,B}$ & $0.04-0.18$ & $0.14-0.44$ & $0.14-0.64$\\
\hline
${\rm Case ~C,D}$ & $0.05-0.19$ & $0.09-0.39$ & $0.45-0.90$\\
\hline
\end{tabular}
\end{center}
\end{table}
\noindent

\begin{table}[H]
\begin{center}
\caption{Prediction of the ranges of $|A_{\mu e}|$ in T2K, No$\nu$A, DUNE} \label{cpas2}
\begin{tabular}{|c|c|c|c|}
\hline
 \multicolumn{4}{|c|}{\cellcolor{gray!30}\textbf{T2K}}\\
\cline{1-4}
\multicolumn{1}{|c|}{${\rm Energy}$}&${E=0.5\hspace{1mm}{\rm GeV}}$&${E=1.0\hspace{1mm}{\rm GeV}}$ &${E=2.0\hspace{1mm}{\rm GeV}}$\\
\cline{1-4}
\multicolumn{1}{|c|}{$\pbox{20cm}{{\rm Case ~A,B}}$} & $\pbox{20cm}{0.14-0.37}$ & $\pbox{20cm}{0.07-0.21}$ &$\pbox{20cm}{0.05-0.10}$\\
\cline{1-4}
\multicolumn{1}{|c|}{$\pbox{20cm}{{\rm Case ~C,D}}$} & $\pbox{20cm}{0.14-0.37}$ & $\pbox{20cm}{0.06-0.19}$ &$\pbox{20cm}{0.05-0.10}$\\
\hline
\cline{1-4}
\multicolumn{4}{|c|}{\cellcolor{gray!30}\textbf{NO$\nu$A}}\\
\cline{1-4}
\multicolumn{1}{|c|}{${\rm Energy}$}&${E=0.5\hspace{1mm}{\rm GeV}}$&${E=1.0\hspace{1mm}{\rm GeV}}$ &${E=2.0\hspace{1mm}{\rm GeV}}$\\
\cline{1-4}
\multicolumn{1}{|c|}{$\pbox{20cm}{{\rm Case ~A,B}}$} & $\pbox{20cm}{0.31-0.80}$ & $\pbox{20cm}{0.21-0.43}$ &$\pbox{20cm}{0.08-0.24}$\\
\cline{1-4}
\multicolumn{1}{|c|}{$\pbox{20cm}{{\rm Case ~C,D}}$} & $\pbox{20cm}{0.29-0.79}$ & $\pbox{20cm}{0.10-0.38}$ &$\pbox{20cm}{0.13-0.29}$\\
\hline
\cline{1-4}
 \multicolumn{4}{|c|}{\cellcolor{gray!30}\textbf{DUNE}}\\
\cline{1-4}
\multicolumn{1}{|c|}{${\rm Energy}$}&${E=0.5\hspace{1mm}{\rm GeV}}$&${E=1.0\hspace{1mm}{\rm GeV}}$ &${E=2.0\hspace{1mm}{\rm GeV}}$\\
\cline{1-4}
\multicolumn{1}{|c|}{$\pbox{20cm}{{\rm Case ~A,B}}$} & $\pbox{20cm}{0.39-0.98}$ & $\pbox{20cm}{0.21-0.64}$ &$\pbox{20cm}{0.15-0.30}$\\
\cline{1-4}
\multicolumn{1}{|c|}{$\pbox{20cm}{{\rm Case ~C,D}}$} & $\pbox{20cm}{0.41-0.97}$ & $\pbox{20cm}{0.61-0.87}$ &$\pbox{20cm}{0.13-0.32}$\\
\cline{1-4}
\end{tabular}
\end{center}
\end{table}
\noindent

\textbf{Flavor flux ratios at neutrino telescopes}- In order to discuss our predictions on the flavor flux ratios at neutrino telescopes (such as IceCube) we deem it necessary to first give a short review of the subject. The main source of ultra high energy cosmic neutrinos are $pp$ and $p\gamma$ collisions\cite{Gandhi:1995tf}. In  $pp$ collisions, protons of TeV$-$PeV range produce neutrinos via the processes $\pi^+\to \mu^+\nu_\mu, \pi^-\to \mu^-\bar{\nu}_\mu, \mu^{+}\to e^+\nu_e\bar{\nu}_\mu$ and $\mu^-\to e^{-}\bar{\nu}_e\nu_\mu.$ Therefore, the normalized flux distributions over different flavors are \begin{equation}\{\phi^S_{\nu_e},\phi^S_{\bar{\nu}_e},\phi^S_{\nu_\mu},\phi^S_{\bar{\nu}_\mu},\phi^S_{\nu_\tau},\phi^S_{\bar{\nu}_\tau}\}=\phi_0\Big\{\frac{1}{6},\frac{1}{6},\frac{1}{3},\frac{1}{3},0,0\Big\},\end{equation} where the superscript $S$ denotes `source' and $\phi_0$ denotes the overall flux normalization. For $p\gamma$ collisions, one is dealing with relatively less energetic $\gamma$-rays (GeV$-$ $10^2$ GeV range). Therefore, the centre-of-mass energy of the $\gamma p$ system can barely allow the reactions $\gamma p\to \Delta^+\to \pi^+ n$ and $\pi^+\to \mu^+\nu_\mu, \mu^+\to e^+\nu_e\bar{\nu}_\mu.$ The corresponding normalized flux distributions over flavor are\begin{equation}\{
\phi^S_{\nu_e},\phi^S_{\bar{\nu}_e},\phi^S_{\nu_\mu},\phi^S_{\bar{\nu}_\mu},\phi^S_{\nu_\tau},\phi^S_{\bar{\nu}_\tau}\}=\phi_0\Big\{\frac{1}{3},0,\frac{1}{3},\frac{1}{3},0,0\Big\}.\end{equation} In either case, if we take $\phi^S_l=\phi^S_{\nu_l}+\phi^S_{\bar{\nu}_l}$ with $l=e,\mu,\tau$, \begin{equation}
\{\phi_e^S,\phi_\mu^S,\phi_\tau\}=\phi_0\Big\{\frac{1}{3},\frac{2}{3},0\Big\}.\label{ratio}
\end{equation}
\noindent

Since neutrino oscillations will change flavor distributions from source (S) to telescope (T)\cite{Xing:2006uk}, the flux reaching the telescope is given by\begin{equation}
\phi^T_{l}\equiv\phi^T_{\nu_l}+\phi^T_{\bar{\nu}_l}=\sum\limits_{m}
\Big[\phi^S_{\nu_m} P(\nu_m\to\nu_l)+\phi^S_{\bar{\nu}_m}P(\bar{\nu}_m\to
\bar{\nu}_l)\Big].\end{equation} Given that the source-to-telescope distance is much greater than the oscillation length, the flavor oscillation probability can be averaged over many oscillations. Hence we have \begin{equation}P(\nu_m\to\nu_l)=P(\bar{\nu}_m\to
\bar{\nu}_l)\approx \sum\limits_{i}|U_{l i}|^2|U_{m i}|^2.\end{equation}
Thus the flux reaching the telescope, after using \eqref{ratio}, will be \begin{equation}
\phi_l^T=\sum\limits_{i}\sum\limits_{m}\phi_m^S|U_{l i}|^2|U_{m i}|^2=\frac{\phi_0}{3}\sum\limits_{i}|U_{l i}|^2(|U_{ei}|^2+2|U_{\mu i}|^2).
\end{equation} Using the unitarity of the PMNS matrix i.e., $|U_{ei}|^2+|U_{\mu i}|^2+|U_{\tau i}|^2=1$, we have \begin{equation}\phi_l^T=\frac{\phi_0}{3}[1+\sum\limits_{i}|U_{l i}|^2(|U_{\mu i}|^2-|U_{\tau i}|^2)]=\frac{\phi_0}{3}[1+\sum\limits_{i}|U_{l i}|^2\Delta_i].\end{equation} where $\Delta_i=|U_{\mu i}|^2-|U_{\tau i}|^2$. If there is exact CP transformed $\mu\tau$ (anti)symmetry, $\Delta_i=0$, and $\phi_e^T=\phi_\mu^T=\phi_\tau^T.$
\noindent

\begin{figure}[H]
\includegraphics[scale=.15]{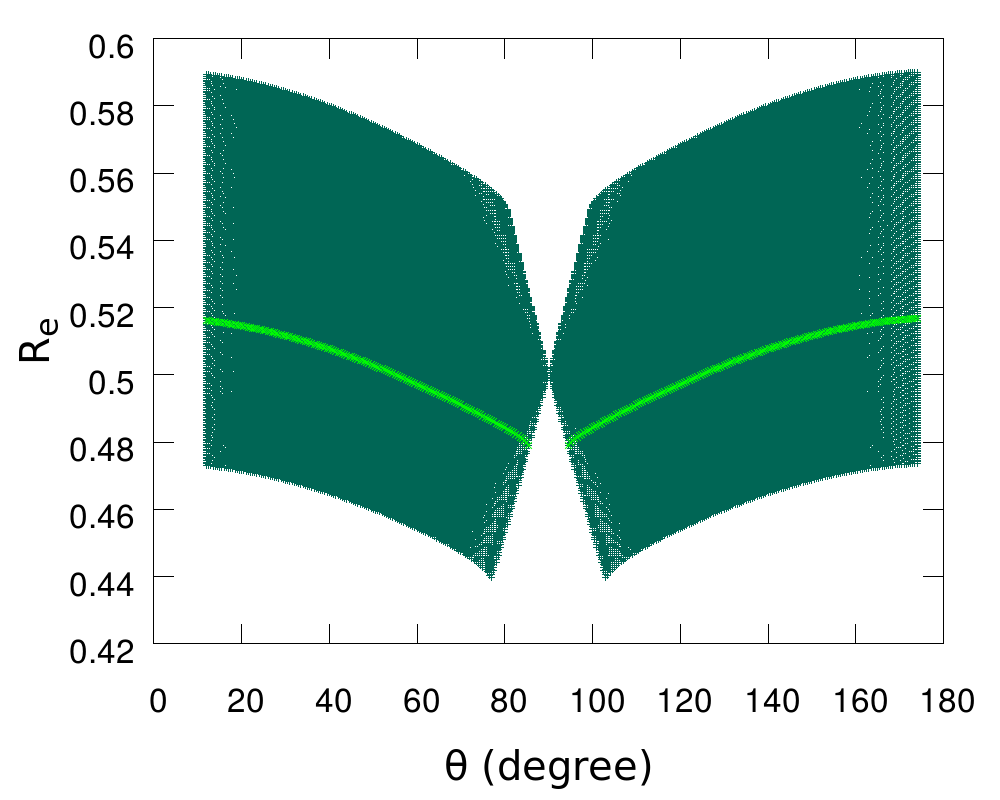}\includegraphics[scale=.15]{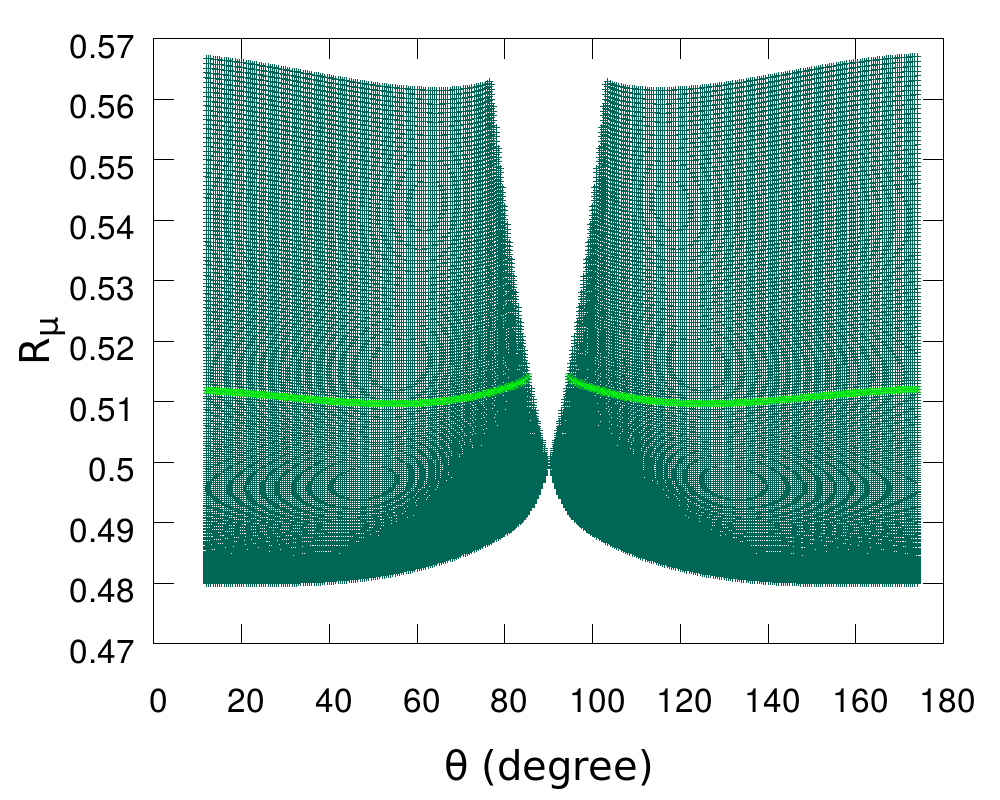}\includegraphics[scale=.15]{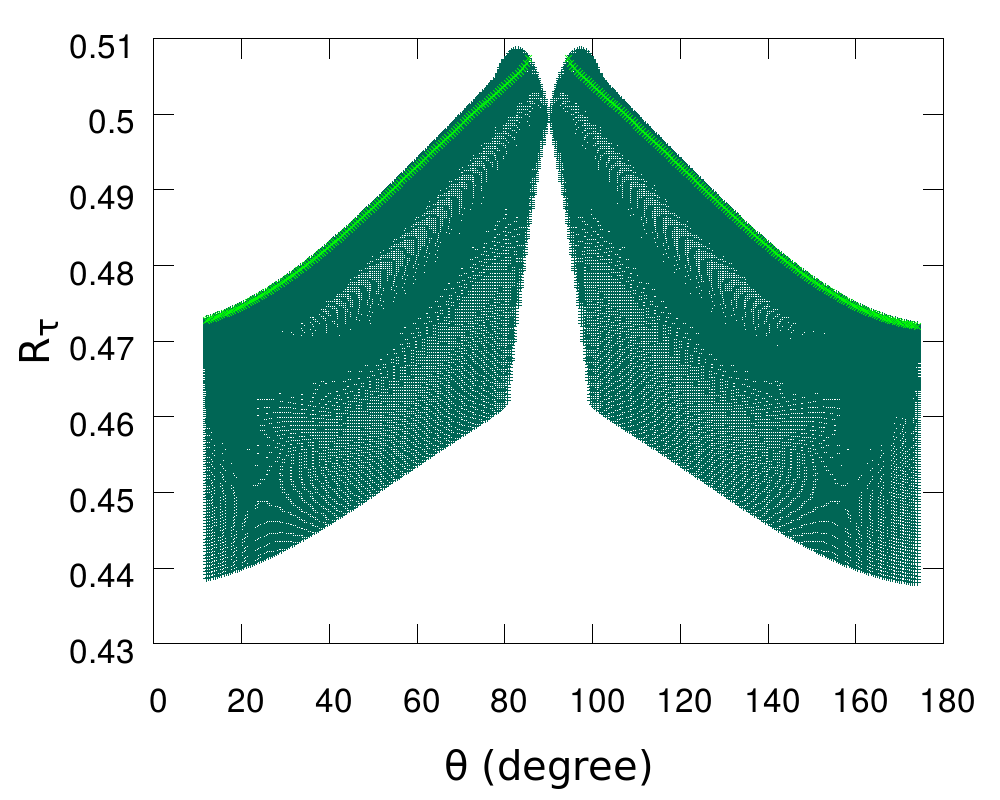}

\includegraphics[scale=.15]{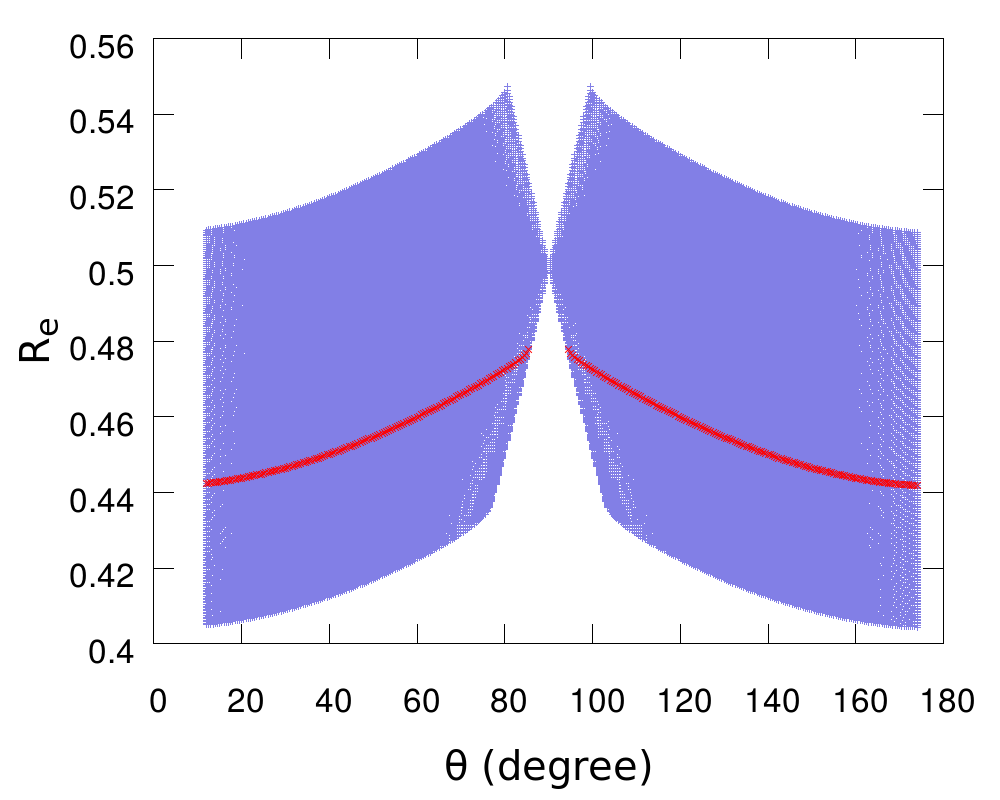}\includegraphics[scale=.15]{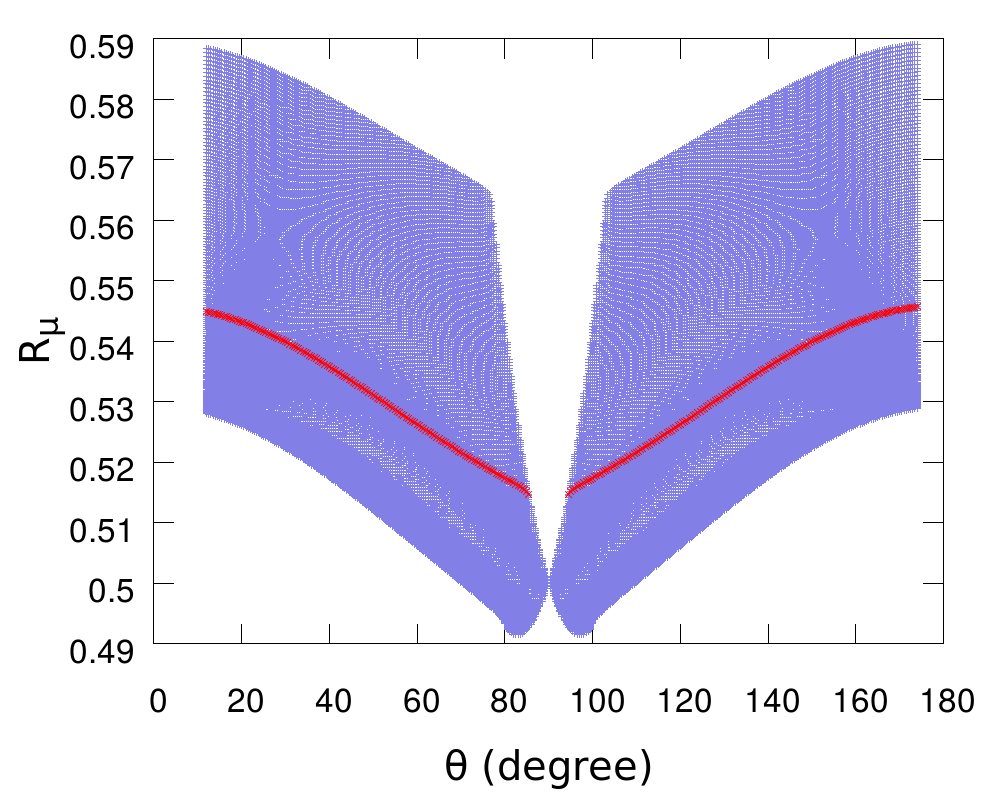}\includegraphics[scale=.15]{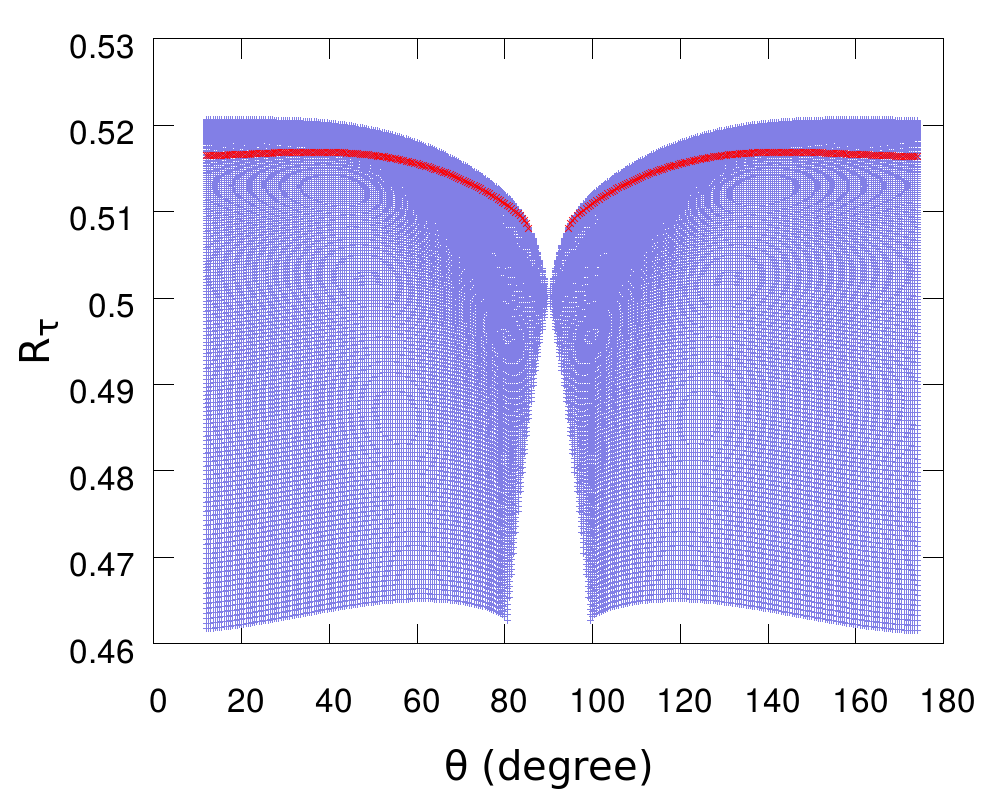}
\caption{Flux ratios $R_e, R_\mu, R_{\tau}$ vs. the $\mu\tau$-mixing parameter $\theta$ for normal ordering. where the three mixing angles have been allowed to vary over their $3\sigma$ ranges. The green(red) line in each plot of the upper(lower) panel corresponds to the best fit value of the mixing angles. The plots in the upper (lower) panel correspond to $\cos\delta\geq 0(\leq 0)$.}\label{figx}
\end{figure}

\begin{figure}[H]
\includegraphics[scale=.15]{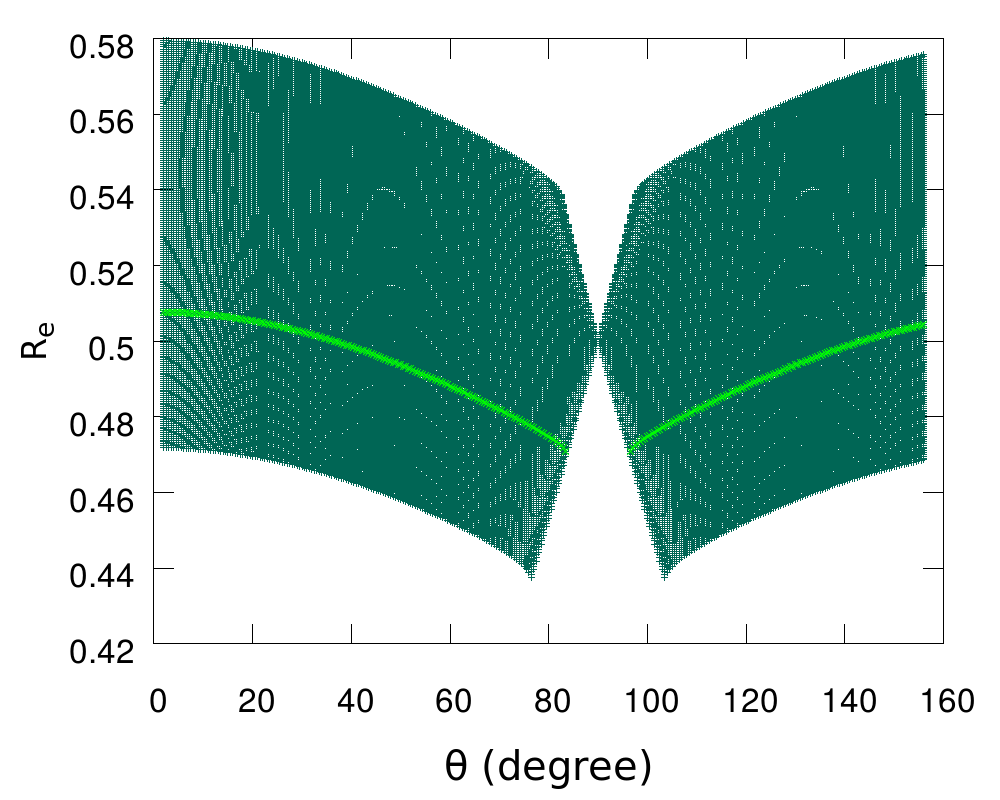}\includegraphics[scale=.15]{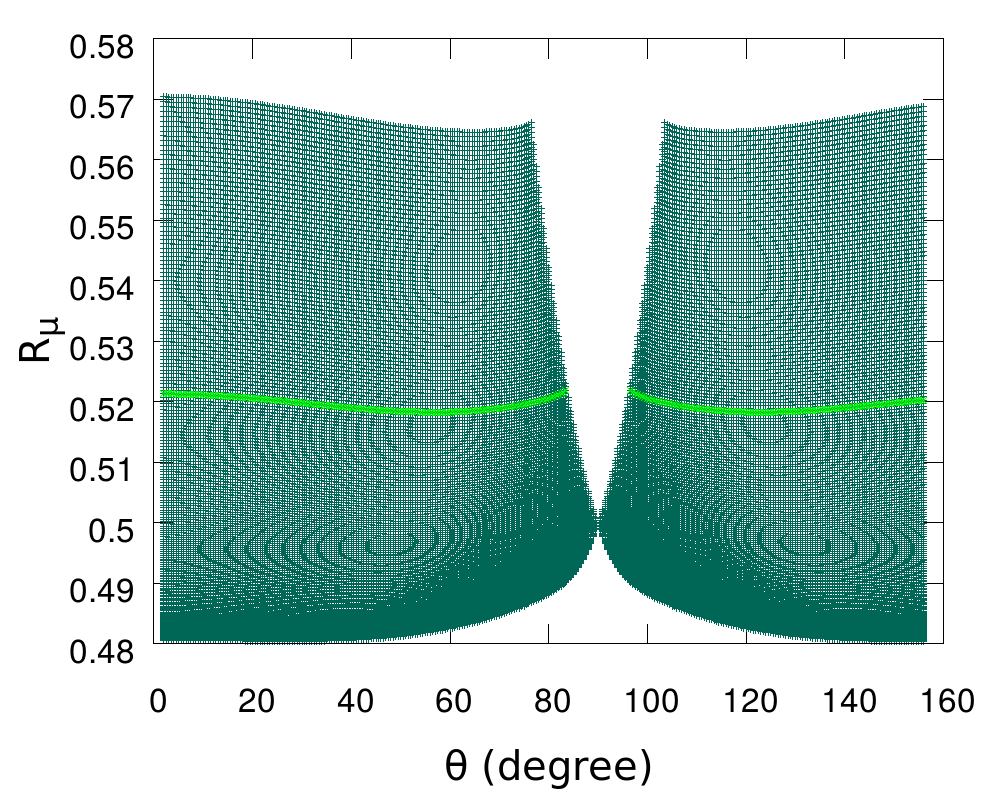}\includegraphics[scale=.15]{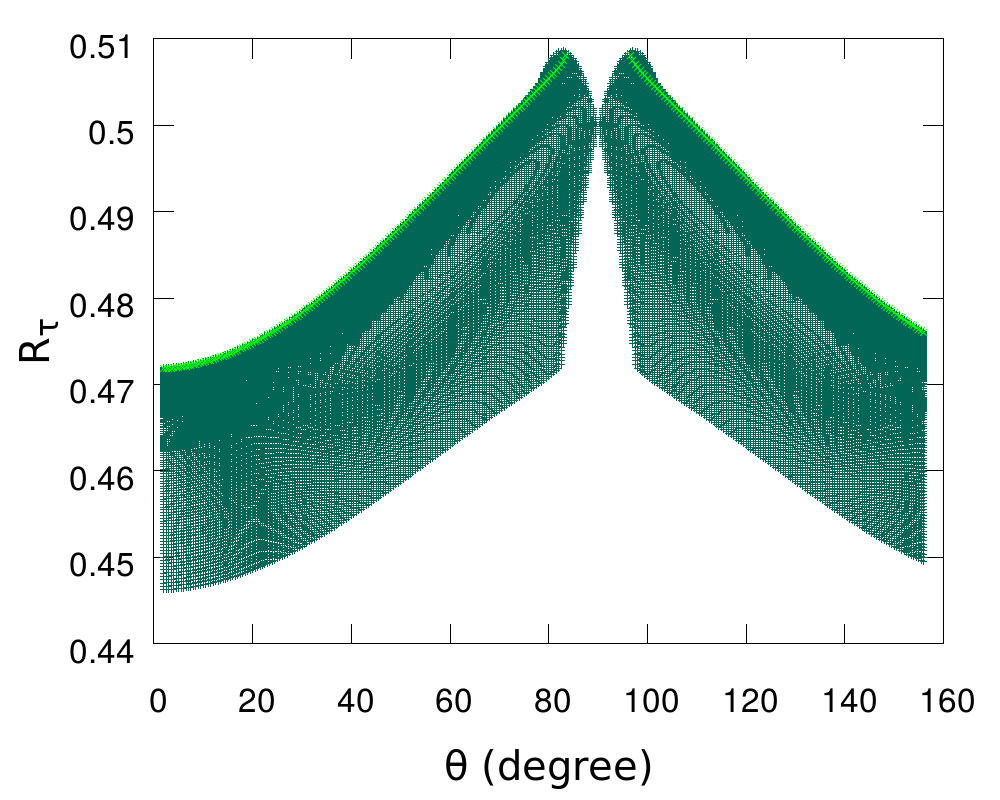}

\includegraphics[scale=.15]{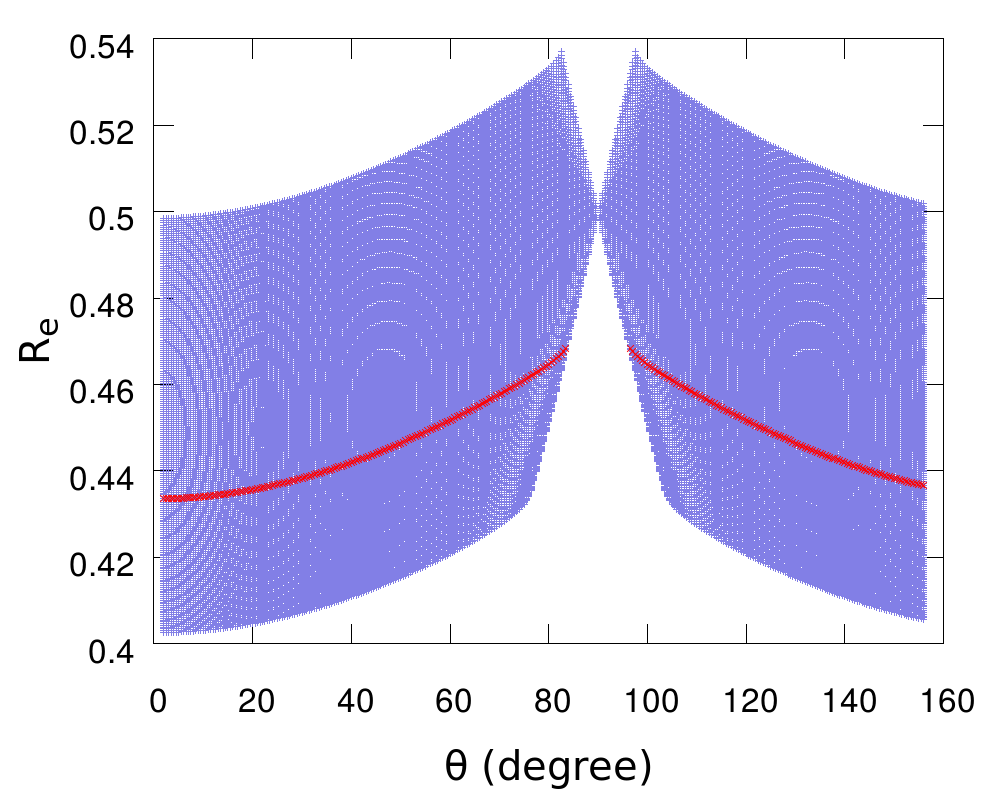}\includegraphics[scale=.15]{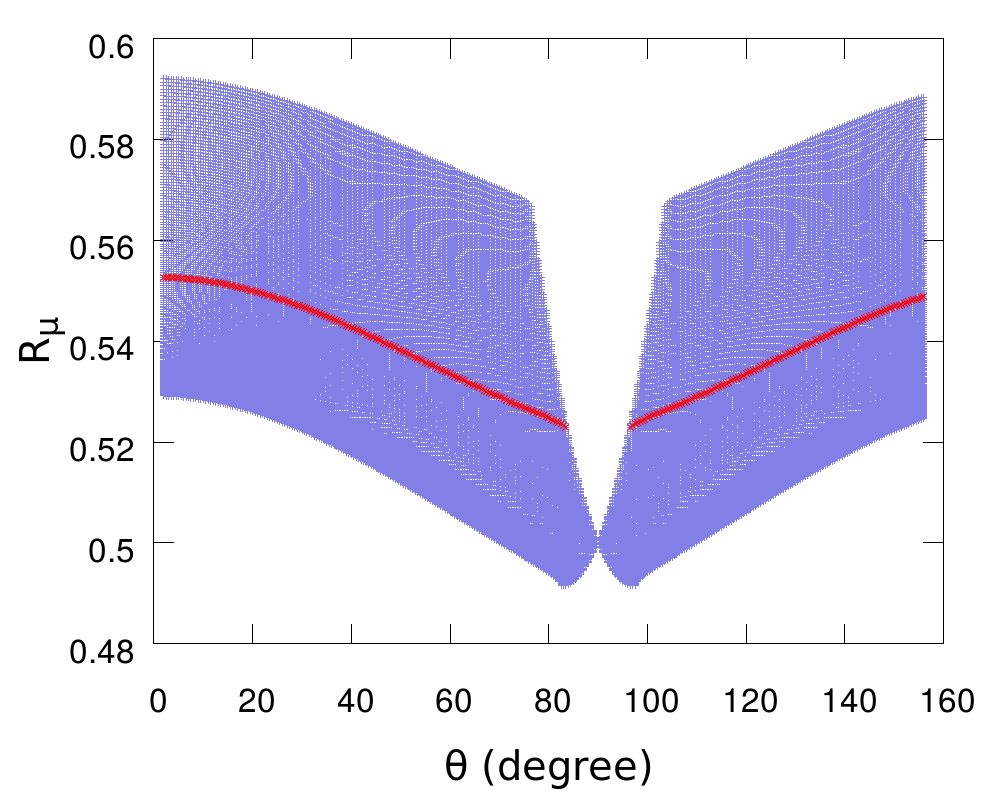}\includegraphics[scale=.15]{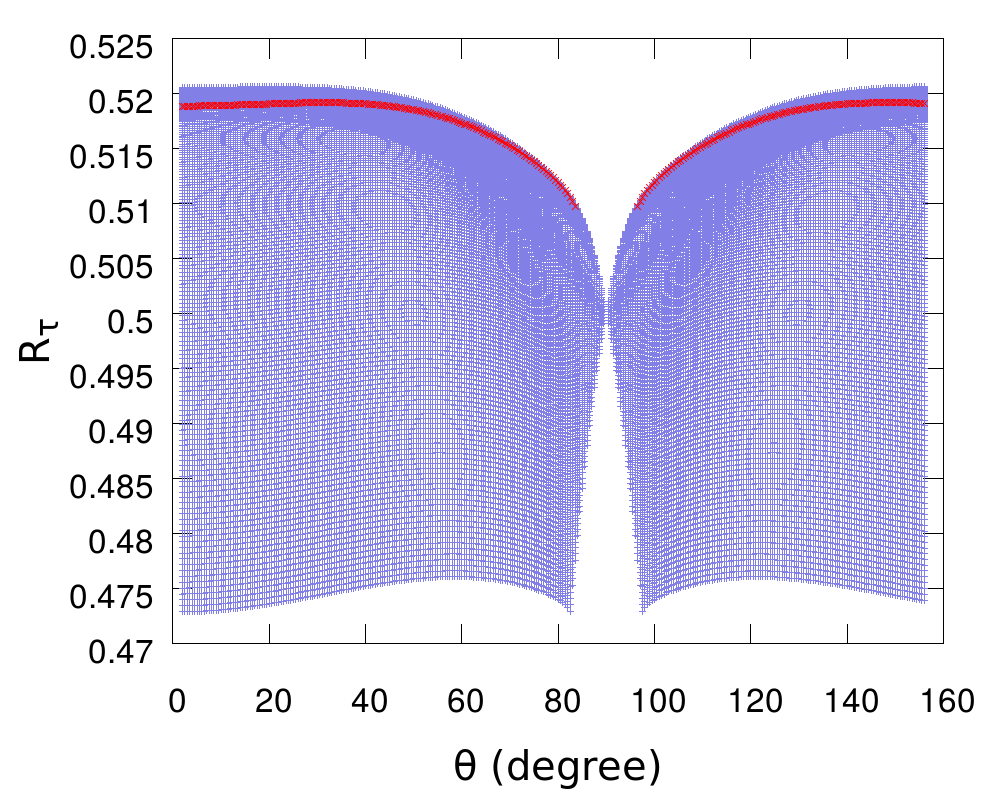}
\caption{Flux ratios $R_e, R_\mu, R_{\tau}$ vs. the $\mu\tau$-mixing parameter $\theta$ for inverted ordering where the three mixing angles have been allowed to vary over their $3\sigma$ ranges. The green(red) line in each plot of the upper(lower) panel corresponds to the best fit value of the mixing angles. The plots in the upper (lower) panel correspond to $\cos\delta\geq 0(\leq 0)$.}\label{figy}
\end{figure}

With the above background, one can define certain flavor flux ratios $R_l$ ($l=e,\mu,\tau$) at the neutrino telescope as
\begin{equation} R_l\equiv\frac{\phi_l^T}{\sum\limits_{m}\phi_m^T-\phi_l^T}=\frac{1+\sum\limits_{i}|U_{l i}|^2\Delta_i}{2-\sum\limits_{i}|U_{l i}|^2\Delta_i},
\end{equation} where $m=e,\mu,\tau$ and $U$ is as in \eqref{eu}. Since $s_{13}^2\approx 0.01$, we can neglect $\mathcal{O}(s^2_{13})$ terms. Then the approximate expressions for the flux ratios become \begin{equation}
R_e\equiv\frac{\phi_e^T}{\phi^T_\mu+\phi^T_\tau}\approx\frac{1+\frac{1}{2}\sin^22\theta_{12}\cos2\theta_{23}+\frac{1}{2}\sin4\theta_{12}\sin2\theta_{23}s_{13}\cos\delta}{2-\frac{1}{2}\sin^22\theta_{12}\cos2\theta_{23}-\frac{1}{2}\sin4\theta_{12}\sin2\theta_{23}s_{13}\cos\delta},\label{a}\end{equation}
\begin{equation}R_\mu\equiv\frac{\phi_\mu^T}{\phi^T_e+\phi^T_\tau}\approx\frac{1+\{c^2_{23}(1-\frac{1}{2}\sin^22\theta_{12})-s^2_{23}\}\cos2\theta_{23}-\frac{1}{4}\sin4\theta_{12}\sin2\theta_{23}s_{13}\cos\delta(4c^2_{23}-1)}{2-\cos^22\theta_{23}+\frac{1}{2}\sin^22\theta_{12}\cos2\theta_{23}c^2_{23}+\frac{1}{4}(3-4s^2_{23})\sin4\theta_{12}\sin2\theta_{23}s_{13}\cos\delta},\label{b}\end{equation}
\begin{equation}R_\tau\equiv\frac{\phi_\tau^T}{\phi^T_e+\phi^T_\mu}\approx
\frac{1+\{s^2_{23}(1-\frac{1}{2}\sin^22\theta_{12})-c^2_{23}\}\cos2\theta_{23}-\frac{1}{4}\sin4\theta_{12}\sin2\theta_{23}s_{13}\cos\delta(4s^2_{23}-1)}{2+\cos^22\theta_{23}+\frac{1}{2}\sin^22\theta_{12}\cos2\theta_{23}c^2_{23}+\frac{1}{4}(3-4c^2_{23})\sin4\theta_{12}\sin2\theta_{23}s_{13}\cos\delta}.\label{c}
\end{equation}
\noindent
Note that each $R_l$ depends on $\cos\delta$ which from \eqref{xop} is given by \begin{equation}
\cos\delta=\pm(\sqrt{\cos^2\theta\sin^22\theta_{23}-\sin^2\theta\cos^22\theta_{23}})
/\sin2\theta_{23}.\end{equation} With $\theta=\pi/2+\epsilon$ for any arbitrary $\epsilon$, positive or negative, we can write \begin{equation}
\cos\delta=\pm(\sqrt{\sin^2\epsilon\sin^22\theta_{23}-\cos^2\epsilon\cos^22\theta_{23}})
/\sin2\theta_{23}\label{top}\end{equation} which is the same whether $\epsilon$ is positive or negative. For either sign, this explains why each $R_l$ in Fig.\ref{figx} and \ref{figy} is symmetric about $\theta=\pi/2$ though the allowed range of $\theta$ is not (Table \ref{osc2}).  The `$\pm$' sign in \eqref{top} tells us that for a fixed $\theta$ (equivalently, for a fixed $\epsilon$), and fixed $\theta_{23}$, each $R_l$ is double-valued except for $\theta=\pi/2$ (i.e., $\epsilon=0$) where $\cos\delta=0$ from \eqref{top} and \eqref{xop}. However, instead of two discrete values of $R_l$, a continuous band is obtained for a fixed $\theta$ since $\theta_{23}$ has been allowed to vary in its current $3\sigma$ range while the other mixing angles are held fixed a their best fit values. In the figure below, we plot the variation of the flavor flux ratios $R_{l}$ with the $\mu\tau$ mixing parameter $\theta$ in its allowed range for both normal and inverted types of mass ordering. Unlike the CP asymmetry parameter in neutrino oscillation experiments, these flavor flux ratios are different for NO and IO$-$specifically in the allowed ranges of $\theta$.  An exact CP transformed $\mu\tau$ interchange (${\rm CP}^{\mu\tau}$) antisymmetry leads to $R_e=R_\mu=R_\tau=1/2$ irrespective of the mass ordering. This can be clearly seen from the approximate expressions of flux ratios in \eqref{a}, \eqref{b} and \eqref{c} in the limit $\theta=\pi/2$ or equivalently, $\theta_{23}=\pi/4$ and $\cos\delta=0$. But a small deviation from ${\rm CP}^{\mu\tau}$ (anti)symmetry may lead to a drastic change of the flux ratios as is clear from the sharp edges of the allowed parameter spaces on either side of $\theta=\pi/2$.

\noindent

In order to obtain precise predictions for flavor flux ratios, a precise value of $\theta$ must be specified. In particular, precise measurements of $\delta$ and $\theta_{23}$ can be used to pinpoint a value of $\theta$ from Eq.\eqref{xop}. As an illustration, the best fit value of $\delta=234^0$ $(278^\circ)$ and $\theta_{23}=47.2^\circ$ $(48.1^\circ)$ for NO (IO), the value of $\theta$ turns out to be $34.75^\circ(75.9^\circ)$. The contours corresponding to the best fit values of the mixing angles has now been indicated in Fig.\ref{figx} and Fig.\ref{figy}. Now, it can be clearly seen that, as $\theta$ deviates from $\pi/2$, the flavor flux ratios deviate drastically from 0.5 and the corresponding values have been tabulated in Table 8. The quantitative predictions of flux ratio $\theta$ deviating from $\pi/2$ has now been summarized in Table \ref{fluxo} the current best fit values $215^\circ(284^\circ)$ of $\delta$ and $49.6^\circ$ $(49.8^\circ)$ of $\theta_{23}$ to obtain $\theta$ to be $34.75^\circ(75.9^\circ)$ for NO(IO) case. The corresponding values of $R_{e},R_\mu$ and $R_\tau$ have been found to be $0.456$ $(0.465)$, $0.529$ $(0.525)$ and $0.516$ $(0.512)$ respectively. It is interesting to note that while the predicted value of $R_e$ is less than $0.5$ those of $R_\mu$ and $R_\tau$ are greater than $0.5$. If this best fit values change in future, the corresponding predictions for $R_l$ can be easily obtained using the formulae 4.16 and 4.20 to test or falsify our proposal.

\noindent

\begin{table}[H]
\begin{center}
\caption{Prediction for the values of flux ratios ($R_l$) for $\theta\neq\pi/2$\cite{Esteban:2016qun}} \label{fluxo}
 \begin{tabular}{|c|c|c|c|c|c|c|c|}
\hline
${\rm Ordering}\downarrow$ & $\text{ bf value of}~\delta$ & $\text{bf value of}~ \theta_{23}$ & $\theta$ & $R_e$ & $R_\mu$ & $R_\tau$\\
\hline
${\rm NO}$ & $234^\circ$ & $47.2^\circ$ & $53.70^\circ$ & $0.456$ & $0.529$ & $0.516$\\
\hline
${\rm IO}$ & $278^\circ$ & $48.1^\circ$ & $79.74^\circ$& $0.465$ & $0.525$ & $0.512$\\
\hline
\end{tabular}
\end{center}
\end{table}

\section{Summary and conclusions}\label{sec5} We have proposed a CP transformed mixed $\mu\tau$ antisymmetry in the light neutrino Majorana mass matrix $M_\nu$ implemented in the Lagrangian by a generalized CP transformation on left-chiral flavor neutrino fields. We explore its consequences in leptonic CP violation. The Dirac CP phase $\delta$, which is in general nonmaximal, is found to be correlated with both the $\mu\tau$ mixing parameter $\theta$ and the atmospheric mixing angle $\theta_{23}$. For a nonmaximal $\delta$, one of the Majorana phases is neither zero nor $\pi$, thereby leading to a nonvanishing Majorana CP violation. Moreover, we discuss the consequences of our proposal on the $\beta\beta0\nu$ decay process in relation to ongoing and upcoming experiments. We have also investigated the quantitative variation of the CP asymmetry parameter $A_{\mu e}$ as a function of beam energy for different baseline lengths as appropriate for different experiments. We have further obtained the implications of $\mu\tau$ mixing on flavor flux ratios $R_{e,\mu,\tau}$ at neutrino telescopes such as IceCube. While an exact $\mu\tau$ interchange antisymmetry leads to $R_e=R_\mu=R_\tau=0.5$, any tiny departure will cause a significant deviation in the flux ratios, as has been explained quantitatively. Further, a careful measurement of these flux ratios in future can put additional constraints on the parameter $\theta$.

\section*{Acknowledgement} We thank R. Samanta for useful discussions. The work of R. Sinha and A. Ghosal is supported by the Department of Atomic Energy (DAE), Government of India. The work of P. Roy has been supported by the Indian National Science Academy.


\begin{thebibliography}{9}
\bibitem{King:2015aea}
  S.~F.~King,
  J.\ Phys.\ G {\bf 42}, 123001 (2015).
\bibitem{Aghanim:2016yuo}
  N.~Aghanim {\it et al.} [Planck Collaboration],
  Astron.\ Astrophys.\  {\bf 596}, A107 (2016).
\bibitem{Abe:2017bay}
K.~Abe {\it et al.} [T2K Collaboration],
  Phys.\ Rev.\ Lett.\  {\bf 118}, no. 15, 151801 (2017).
\bibitem{Adamson:2017qqn}
  P.~Adamson {\it et al.} [NOvA Collaboration],
  Phys.\ Rev.\ Lett.\  {\bf 118}, no. 15, 151802 (2017).
 P.~Adamson {\it et al.} [NOvA Collaboration],
  Phys.\ Rev.\ Lett.\  {\bf 118}, no. 23, 231801 (2017).
 A. Himmel (NOvA),
New neutrino oscillation results from NOVA
,\url{https://indico.cern.ch/event/696410/
(2018)}.
\bibitem{minos}
P.~Adamson {\it et al.} [MINOS Collaboration],
  Phys.\ Rev.\ Lett.\  {\bf 110}, no. 25, 251801 (2013).
    P.~Adamson {\it et al.} [MINOS Collaboration],
  Phys.\ Rev.\ Lett.\  {\bf 110}, no. 17, 171801 (2013).
  \bibitem{reno}
 S.-H. Seo (RENO), in
15th International Conference on Topics in Astroparticle and Underground Physics (TAUP 2017) Sudbury, Ontario, Canada, July 24-28, 2017 (2017), 1710.08204.
\bibitem{Esteban:2016qun}
  I.~Esteban, M.~C.~Gonzalez-Garcia, M.~Maltoni, I.~Martinez-Soler and T.~Schwetz,
  JHEP {\bf 1701}, 087 (2017).


 \bibitem{Grimus:2003yn}
   P.~F.~Harrison and W.~G.~Scott,
  Phys.\ Lett.\ B {\bf 547}, 219 (2002).
  W.~Grimus and L.~Lavoura,
  Phys.\ Lett.\ B {\bf 579}, 113 (2004).
 E.~Ma,
  Phys.\ Lett.\ B {\bf 752}, 198 (2016).
   R.~N.~Mohapatra and C.~C.~Nishi,
  Phys.\ Rev.\ D {\bf 86}, 073007 (2012).
  R.~N.~Mohapatra and C.~C.~Nishi,
  JHEP {\bf 1508}, 092 (2015).
 S.~F.~Ge, D.~A.~Dicus and W.~W.~Repko,
  Phys.\ Lett.\ B {\bf 702}, 220 (2011).
  S.~F.~Ge, D.~A.~Dicus and W.~W.~Repko,
  Phys.\ Rev.\ Lett.\  {\bf 108}, 041801 (2012).
  R.~Samanta, P.~Roy and A.~Ghosal,
  Eur.\ Phys.\ J.\ C {\bf 76}, no. 12, 662 (2016).
  R.~Sinha, R.~Samanta and A.~Ghosal,
  JHEP {\bf 1712}, 030 (2017).

\bibitem{Samanta:2018efa}
  R.~Samanta, R.~Sinha and A.~Ghosal,
  arXiv:1805.10031 [hep-ph].
W.~Grimus, A.~S.~Joshipura, S.~Kaneko, L.~Lavoura, H.~Sawanaka and M.~Tanimoto,
  Nucl.\ Phys.\ B {\bf 713}, 151 (2005).




\bibitem{Ishimori:2010au}
  G.~Altarelli and F.~Feruglio,
  Rev.\ Mod.\ Phys.\  {\bf 82}, 2701 (2010).
   H.~Ishimori, T.~Kobayashi, H.~Ohki, Y.~Shimizu, H.~Okada and M.~Tanimoto,
  Prog.\ Theor.\ Phys.\ Suppl.\  {\bf 183}, 1 (2010).
  S.~F.~King,
  Prog.\ Part.\ Nucl.\ Phys.\  {\bf 94}, 217 (2017).
  S.~T.~Petcov,
  Eur.\ Phys.\ J.\ C {\bf 78}, no. 9, 709 (2018).
  S.~T.~Petcov,
  Nucl.\ Phys.\ B {\bf 892} (2015) 400.


\bibitem{mutaus} 
R.~N.~Mohapatra and S.~Nussinov,
  Phys.\ Rev.\ D {\bf 60}, 013002 (1999).
  T.~Fukuyama and H.~Nishiura,
  hep-ph/9702253.
 C.~S.~Lam,
  Phys.\ Lett.\ B {\bf 507}, 214 (2001).
  E.~Ma and M.~Raidal,
  Phys.\ Rev.\ Lett.\  {\bf 87}, 011802 (2001)
  Erratum: [Phys.\ Rev.\ Lett.\  {\bf 87}, 159901(E) (2001)].
  K.~R.~S.~Balaji, W.~Grimus and T.~Schwetz,
  Phys.\ Lett.\ B {\bf 508}, 301 (2001).
     A.~Ghosal,
  Mod.\ Phys.\ Lett.\ A {\bf 19}, 2579 (2004).
  Z.~z.~Xing and Z.~h.~Zhao,
  Rept.\ Prog.\ Phys.\  {\bf 79}, no. 7, 076201 (2016).
  S.~F.~Ge, H.~J.~He and F.~R.~Yin,
  JCAP {\bf 1005}, 017 (2010).

\bibitem{An:2015rpe}
  F.~P.~An {\it et al.} [Daya Bay Collaboration],
  Phys.\ Rev.\ Lett.\  {\bf 115}, no. 11, 111802 (2015).


\bibitem{CPt}
 G. Ecker, W. Grimus, H. Neufeld, J.Phys. A20, L807 (1987).
  W.~Grimus and M.~N.~Rebelo,
  Phys.\ Rept.\  {\bf 281}, 239 (1997).
 R.~N.~Mohapatra and C.~C.~Nishi,
  Phys.\ Rev.\ D {\bf 86}, 073007 (2012).
 S.~Gupta, A.~S.~Joshipura and K.~M.~Patel,
  Phys.\ Rev.\ D {\bf 85}, 031903 (2012).
F.~Feruglio, C.~Hagedorn and R.~Ziegler,
  JHEP {\bf 1307}, 027 (2013).
  M.~Holthausen, M.~Lindner and M.~A.~Schmidt,
  JHEP {\bf 1304}, 122 (2013).
   M.~C.~Chen, M.~Fallbacher, K.~T.~Mahanthappa, M.~Ratz and A.~Trautner,
  Nucl.\ Phys.\ B {\bf 883}, 267 (2014).
    G.~J.~Ding, S.~F.~King, C.~Luhn and A.~J.~Stuart,
  JHEP {\bf 1305}, 084 (2013).
  G.~J.~Ding, S.~F.~King and A.~J.~Stuart,
  JHEP {\bf 1312}, 006 (2013).
  G.~J.~Ding and Y.~L.~Zhou,
  JHEP {\bf 1406}, 023 (2014).
  G.~J.~Ding and S.~F.~King,
  Phys.\ Rev.\ D {\bf 89}, no. 9, 093020 (2014).
  C.~C.~Li and G.~J.~Ding,
  Nucl.\ Phys.\ B {\bf 881}, 206 (2014).
  S.~F.~King and T.~Neder,
  Phys.\ Lett.\ B {\bf 736}, 308 (2014).
  C.~C.~Li and G.~J.~Ding,
  JHEP {\bf 1505}, 100 (2015).
  G.~J.~Ding and Y.~L.~Zhou,
  Chin.\ Phys.\ C {\bf 39}, no. 2, 021001 (2015).
  F.~Feruglio, C.~Hagedorn and R.~Ziegler,
  Eur.\ Phys.\ J.\ C {\bf 74}, 2753 (2014).
  A.~Di Iura, C.~Hagedorn and D.~Meloni,
  JHEP {\bf 1508}, 037 (2015).
  C.~Hagedorn, A.~Meroni and E.~Molinaro,
  Nucl.\ Phys.\ B {\bf 891}, 499 (2015).
  P.~Chen, C.~Y.~Yao and G.~J.~Ding,
  Phys.\ Rev.\ D {\bf 92}, no. 7, 073002 (2015).
 C.~C.~Nishi,
  Phys.\ Rev.\ D {\bf 93}, no. 9, 093009 (2016).
   C.~C.~Nishi and B.~L.~Sánchez-Vega,
  JHEP {\bf 1701}, 068 (2017).
   W.~Rodejohann and X.~J.~Xu,
  Phys.\ Rev.\ D {\bf 96}, no. 5, 055039 (2017).
  I.~Girardi, A.~Meroni, S.~T.~Petcov and M.~Spinrath,
  JHEP {\bf 1402}, 050 (2014).
  J.~T.~Penedo, S.~T.~Petcov and A.~V.~Titov,
  JHEP {\bf 1712}, 022 (2017).
  L.~L.~Everett, T.~Garon and A.~J.~Stuart,
  JHEP {\bf 1504}, 069 (2015).
  L.~L.~Everett and A.~J.~Stuart,
  Phys.\ Rev.\ D {\bf 96}, no. 3, 035030 (2017).
 A comprehensive review : S.~F.~King,
  Prog.\ Part.\ Nucl.\ Phys.\  {\bf 94}, 217 (2017).



\bibitem{Chen:2015siy}
  P.~Chen, G.~J.~Ding, F.~Gonzalez-Canales and J.~W.~F.~Valle,
  Phys.\ Lett.\ B {\bf 753}, 644 (2016).


\bibitem{newadd}
  S.~F.~King and C.~C.~Nishi,
  Phys.\ Lett.\ B {\bf 785}, 391 (2018).
  S.~F.~King and Y.~L.~Zhou,
  arXiv:1901.06877 [hep-ph].


\bibitem{Samanta:2017kce}
  R.~Samanta, P.~Roy and A.~Ghosal,
  JHEP {\bf 1806} (2018) 085.



\bibitem{Grimus}
  W.~Grimus, S.~Kaneko, L.~Lavoura, H.~Sawanaka and M.~Tanimoto,
  JHEP {\bf 0601}, 110 (2006)



\bibitem{joshi}

A.~S.~Joshipura,
JHEP {\bf 1511}, 186 (2015).
A.~S.~Joshipura and N.~Nath,
Phys.\ Rev.\ D {\bf 94}, no. 3, 036008 (2016).



  \bibitem{Joshipura:2015dsa}
  A.~S.~Joshipura and K.~M.~Patel,
  Phys.\ Lett.\ B {\bf 749}, 159 (2015).

  \bibitem{Tanabashi:2018oca}
  M.~Tanabashi {\it et al.} [ParticleDataGroup],
  Phys.\ Rev.\ D {\bf 98}, no. 3, 030001 (2018).




\bibitem{Bernabeu:1986fc}
  J.~Bernabeu, G.~C.~Branco and M.~Gronau,
  Phys.\ Lett.\  {\bf 169B}, 243 (1986).
  G.~C.~Branco, L.~Lavoura and M.~N.~Rebelo,
  Phys.\ Lett.\ B {\bf 180}, 264 (1986).

\bibitem{Branco}
   G. C. Branco, L. Lavoura and J. P. Silva, {\bf CP Violation}, (Clarendon Press, Oxford, 1999)
  J.~Iizuka, Y.~Kaneko, T.~Kitabayashi, N.~Koizumi and M.~Yasue,
  Phys.\ Lett.\ B {\bf 732}, 191 (2014).
  R.~Samanta and A.~Ghosal,
  Nucl.\ Phys.\ B {\bf 911}, 846 (2016).
R.~Samanta, M.~Chakraborty and A.~Ghosal,
  Nucl.\ Phys.\ B {\bf 904}, 86 (2016).






\bibitem{Rahat:2018sgs}
  M.~H.~Rahat, P.~Ramond and B.~Xu,
  arXiv:1805.10684. [hep-ph].


\bibitem{Adhikary:2013bma}
  B.~Adhikary, M.~Chakraborty and A.~Ghosal,
  JHEP {\bf 1310}, 043 (2013)
  Erratum: [JHEP {\bf 1409}, 180 (2014)].


\bibitem{Rodejohann:2011mu}
  W.~Rodejohann,
  Int.\ J.\ Mod.\ Phys.\ E {\bf 20}, 1833 (2011).
  P.~S.~Bhupal Dev, S.~Goswami, M.~Mitra and W.~Rodejohann,
  Phys.\ Rev.\ D {\bf 88}, 091301 (2013).




\bibitem{gerda2}

  M.~Agostini {\it et al.} [GERDA Collaboration],
  Phys.\ Rev.\ Lett.\  {\bf 120}, no. 13, 132503 (2018).

  \bibitem{Agostini:2017jim}
  M.~Agostini, G.~Benato and J.A.~Detwiler,
  Phys.\ Rev.\ D {\bf 96}, no. 5, 053001 (2017).




\bibitem{Gandhi:1995tf}
R.~Gandhi, C.~Quigg, M.~H.~Reno and I.~Sarcevic,
  Astropart.\ Phys.\  {\bf 5}, 81 (1996).
  R.~Gandhi, C.~Quigg, M.~H.~Reno and I.~Sarcevic,
  Phys.\ Rev.\ D {\bf 58}, 093009 (1998).

\bibitem{Xing:2006uk}
Z.~Z.~Xing and S.~Zhou,
Phys.\ Rev.\ D {\bf 74}, 013010 (2006).
  W.~Rodejohann,
  JCAP {\bf 0701}, 029 (2007).
  Z.~z.~Xing,
  Phys.\ Lett.\ B {\bf 716}, 220 (2012).


\end{thebibliography}
\end{document}